\documentclass[10pt,a4paper]{article}
\usepackage{jcappub} 
\usepackage[english]{babel}
\usepackage[T1]{fontenc}
\usepackage[utf8]{inputenc}

\graphicspath{ {./Figures/} }

\usepackage{amsmath, amsfonts, amssymb}
\usepackage{graphicx, color, bm}

\newcommand{\hMpc}{\ h^{-1}\text{Mpc}}
\newcommand{\hMpcsq}{\ h^{-2}\text{Mpc}^2}

\newcommand{\ihMpc}{\ h\text{Mpc}^{-1}}

\newcommand{\barr}{\begin{align}}
\newcommand{\earr}{\end{align}}
\newcommand{\be}{\begin{equation}}
\newcommand{\ee}{\end{equation}}
\newcommand{\eea}{\end{eqnarray}}
\newcommand{\bea}{\begin{eqnarray}}
\newcommand{\eh}[1]{\exp\left[#1\right]}
\newcommand{\la}{\left\langle}
\newcommand{\ra}{\right\rangle}

\newcommand{\ddir}{\delta^\text{(D)}}
\newcommand{\derd}{\text{d}}

\newcommand{\vk}{\vec k}

\renewcommand{\vec}{\bm}

\newcommand{\ii}{\text{i}}
\title{On the reach of perturbative methods for dark matter density fields}
\author[\dagger]{Tobias Baldauf,}
\author[*]{Emmanuel Schaan}
\author[\dagger]{and Matias Zaldarriaga}
\affiliation[\dagger]{School of Natural Sciences, Institute for Advanced Study, Princeton, NJ 08540, U.S.A.}
\affiliation[*]{Department of Astrophysical Sciences, Princeton University, Princeton, NJ 08540, U.S.A.}

\emailAdd{baldauf@ias.edu}
\emailAdd{eschaan@astro.princeton.edu}
\emailAdd{matiasz@ias.edu}

\abstract{We study the mapping from Lagrangian to Eulerian space in the context of the Effective Field Theory (EFT) of Large Scale Structure. 
We compute Lagrangian displacements with Lagrangian Perturbation Theory (LPT) and perform the full non-perturbative transformation from displacement to density. When expanded up to a given order, this transformation reproduces the standard Eulerian Perturbation Theory (SPT) at the same order. However, the full transformation from displacement to density also includes higher order terms. These terms explicitly resum long wavelength motions, thus making the resulting density field better correlated with the true non-linear density field. As a result, the regime of validity of this approach is expected to extend that of the Eulerian EFT, and match that of the IR-resummed Eulerian EFT. This approach thus effectively enables a test of the IR-resummed EFT at the field level.
We estimate the size of stochastic, non-perturbative contributions to the matter density power spectrum.
We find that in our highest order calculation, at redshift $z=0$ the power spectrum of the density field is reproduced with an accuracy of $1\%$ ($10\%$) up to $k=0.25\ihMpc$ ($k=0.46\ihMpc$). We believe that the dominant source of the remaining error is the stochastic contribution. Unfortunately, on these scales the stochastic term does not yet scale as $k^4$ as it does in the very low $k$ regime. Thus, modeling this contribution might be challenging.    
}

\begin{document}

\maketitle 
\section{Introduction}

The Large Scale Structure, i.e., the distribution of matter and galaxies on large scales, has the potential to constrain the history and composition of the Universe in a way that is complementary to the Cosmic Microwave Background (CMB). It even has the potential to tighten CMB constraints on the physics of the early and late accelerated expansion. The latter can be explored through precise measurements of the expansion history using the Baryon Acoustic Oscillation (BAO) \cite{Eisenstein:1998tu} method, while inflation and the generation of the seeds for structure formation can be constrained using primordial non-Gaussianities \cite{Alvarez:2014vva,Dalal:2007cu}. In order to be able to extract the full potential of present and upcoming surveys in answering these fundamental questions, we require accurate predictions for the clustering statistics. On the largest scales linear theory correctly describes the growth of structure, but on smaller scales, non-linear effects become important before on the smallest scales overdensities collapse to form virialized objects. Numerical simulations of structure formation have become a powerful tool to model non-linearities and interpolation techniques have been developed to allow for a fast exploration of the parameter space \cite{Lawrence:2009uk,Heitmann:2013bra}. Yet, numerical simulations hide some of the physics and convergence between codes and parameter choices struggle to pass the percent barrier \cite{Schneider:2015yka,Smith:2012uz}.

Perturbative techniques (for a review see \cite{Bernardeau:2001qr}) are an alternative, extending the validity of linear theory into the weakly non-linear regime and improving precision on large scales. 
Recent years have seen a resurgence of interest in perturbative approaches to study the development of structure in our Universe. A new development has been the introduction of Effective Theory techniques \cite{Baumann:2010tm,Porto:2013qua,Carrasco:2012cv,Mercolli:2013bsa,Pajer:2013jj,Senatore:2014via} and a substantial effort went into higher order computations and testing them against numerical simulations \cite{Carrasco:2013sva,Carrasco:2013mua,Baldauf:2014qfa,Baldauf:2015tla}. 
These approaches have the potential of providing very accurate calculations for observables on large, linear and mildly non-linear scales. 
This work follows our recent paper  \cite{Baldauf:2015tla} (hereafter BSZ), where we studied the perturbative solution for the displacement field of dark matter particles using the appropriate Lagrangian Effective Theory (LEFT) \cite{Porto:2013qua} and compared these results against numerical simulations. 
Tests of LPT at the density field level have also been performed in \cite{Chan:2013vao,Kitaura:2012tj} and tests of Effective Field Theory in connection to LPT have been performed in \cite{McQuinn:2015tva,Vlah:2015sea}.
Here we use the density field induced by those displacements and compare it with the same suite of simulations. We are thus testing the original Eulerian Effective Theory of Large Scale Structure  \cite{Baumann:2010tm}. This paper can also be seen as and extension of \cite{Tassev:2012cq} that we use similar techniques and concepts such as the introduction of transfer functions. Here the focus is on an interpretation of the results in the EFT framework.

As we did in BSZ for the displacements, the goal of this study is to test the range of validity of perturbative approaches at the field level rather than the level of $n$-point statistics. 
We do so by comparing $N$-body results to the perturbative calculation for the same initial conditions, thus avoiding errors arising from sample variance, {\it i.e.} the fact that with the $N$-body code we have only simulated one possible realization of the stochastic initial conditions. This is important because some of the effects we are after are small and are thus difficult to isolate from $n$-point statistics in the presence of sample variance for reasonable size simulation volumes and numbers of realizations. A similar approach was used by \cite{Roth:2011ru} to confront SPT with simulations. 

This paper is structured as follows: we will first review the Lagrangian dynamics and the mapping to densities in Sec.~\ref{sec:dynamics}. We will then compute the density power spectra and fields arising from LPT and compare them to simulations in Sec.~\ref{sec:oneloop}. We then extend this framework to estimate its ultimate reach by allowing for free transfer functions in Sec.~\ref{sec:beyondone}. We will uncover an irreducible error that we associate with the stochastic term and compare this term to the Lagrangian stochastic term in Sec.~\ref{sec:stoch}.
We will conclude and summarize our findings in Sec.~\ref{sec:concl}.

\section{Dynamics in the EFT framework}\label{sec:dynamics}
The  Eulerian position of particles is given by the sum of their initial Lagrangian position $\vec q$ and the subsequent displacement $\vec s (\vec q)$
\be
\vec x=\vec q+\vec s(\vec q)\; .
\ee
The displacement in LEFT is governed by the equation of motion \cite{Porto:2013qua,McQuinn:2015tva,Baldauf:2015tla,Vlah:2015sea}: 
\be
\ddot{\vec s}+\mathcal{H} \dot{\vec s}=-\vec \nabla \phi(x) + \vec a_\text{ct} + \vec a_\text{stoch},
\label{LEFTeq}
\ee
which is solved perturbatively in powers of the density (or its power spectrum) \cite{Bouchet:1994xp}.  The gravitational potential is related to the density via the Poisson equation $\Delta \phi=3/2 \Omega_\text{m} {\cal H}^2 \delta$ and the density is related to the displacement via the mass conservation equation $\left[1+\delta(\vec x)\right]\derd^3 x=\derd^3 q$.

The perturbative expansion is not guaranteed to converge to the correct answer on small scales even if all terms in the series are included (see \cite{McQuinn:2015tva} for an illustration of this in 1 dimension where the perturbative theory can be summed). Furthermore, the mistake on small scales affects the large scales when   statistics at orders higher than tree-level are computed because the momenta in the loops can become large. To fix this problem one needs to modify the dynamics. This is accomplished by adding the additional acceleration terms $\vec a_\text{ct} $ and $\vec a_\text{stoch}$ on the right hand side of equation (\ref{LEFTeq}). The term $\vec a_\text{ct} $ is the part of the additional acceleration which can be explicitly computed in terms of the perturbative solution of the equations. On the other hand only the statistical properties of  $\vec a_\text{stoch}$ can be computed, and they are not correlated with the deterministic part. 

Symmetries dictate the structure of $\vec a_\text{ct}$  and $\vec a_\text{stoch}$. For example in cosmologies similar to our own to compute the one loop power spectrum, the leading correction comes from $\vec a_\text{ct}$ and has the form:  $\vec a_\text{ct}=l^2 (t) \vec \nabla  (\vec \nabla\cdot \vec s)$ where  $l$ has units of length \cite{Porto:2013qua}. It is important to stress that this EFT is only meant to be a good description of the dynamics on large scales. On small scales, smaller than the non-linear scale, the displacements computed using LEFT are not a good approximation to the actual ones.

In the absence of  $\vec a_\text{ct} $ and $\vec a_\text{stoch}$, {\it i.e.} in the standard Lagrangian perturbation theory (LPT), the scalar part of the displacement field can be solved for recursively in terms of the underlying linear density field ($\delta_0$):
\be
\vec s^{(n)}(\vec k)=-\frac{\ii}{n!}\frac{\vec k}{k^2}\prod_{i=1}^n\biggl\{\int_{\vec p_i} \delta_0(\vec p_i)\biggr\}\ddir(\vec p_1+\ldots+\vec p_n)l_n(\vec p_1,\ldots,\vec p_n)\; ,
\ee
where $k=\vec p_1+\ldots +\vec p_n$ and $l_n$ are kernels than can be found in the literature, {\it e.g.} \cite{Bouchet:1994xp,Matsubara:2008re,Zheligovsky:2013eca,Matsubara:2015re}.\\
The density field arising from the displaced particles reads as
\be
\delta(\vec k)=\int \derd^3q \eh{\ii \vec k\cdot \bigl(\vec q+\vec s(\vec q)\bigr)}-(2\pi)^3\ddir(\vec k)\; ,
\label{eq:densmap}
\ee
and the density power spectrum is given by
\be
\label{eq:Pk}
P_\delta(k)=\int \derd^3r \eh{\ii\vec k\cdot \vec r}\Bigl[\la\eh{\ii \vec k \cdot \Delta \vec s}\ra-1\Bigr]\; ,
\ee
where $\Delta \vec s=\vec s(\vec q_2)-\vec s(\vec q_1)$ and $\vec r=\vec q_2-\vec q_1$.
The expectation value of the exponential can be evaluated using the cumulant expansion theorem
\be
\begin{split}
\label{Keq}
\la \eh{-i \vec k \cdot \Delta \vec s}\ra=\exp \Bigl[&-\frac{1}{2}k_i k_j \la \Delta s_i  \Delta s_j\ra-\frac{1}{3!}\ii k_i k_j k_l \la \Delta s_i  \Delta s_j\Delta s_l\ra_\text{c}\\
&+ \frac{1}{4!}k_i k_j k_l k_m \la \Delta s_i  \Delta s_j\Delta s_l\Delta s_m\ra_\text{c}+\ldots\Bigr],
\end{split}
\ee
where the subscript c denotes the connected part of the correlator. 

If one where to expand the exponential in equation (\ref{Keq}) and keep only terms up to a given order in the power spectrum, the resulting series is identical to that of Standard Perturbation Theory (SPT)\footnote{For a review of SPT please see \cite{Bernardeau:2001qr}.}. 
In almost all respects these are the only pieces of (\ref{Keq}) that are trustworthy. The one exception are terms associated with large scale motions which are fixed by the equivalence principle and are resummed correctly to all orders by  (\ref{Keq}). These terms are of interest in practice as they are the ones responsible for the smoothing of the BAO peak. The other terms kept by the exponentiation are a superset of the terms one should keep to be consistent at a given order. As long as these terms are small, keeping them does no harm and we will mostly do so for computational convenience. If we were only to leave exponentiated the bulk motion pieces, one would end up with formulas analog to those in the so-called IR-resummation procedure \cite{Senatore:2014via,Baldauf:2015xfa}. We will see that in practice keeping terms in the exponential is not particularly harmful. 

Even though the trustworthy part of equation  (\ref{Keq}) coincided with what one can compute using SPT with the IR-resummation, we found it useful to be able to include the effects of the bulk motions at higher orders automatically and at the field level. This is so because in this paper we will determine the coefficient in the EFT counter terms by correlating the $N$-body results with the perturbative calculations. If one does this directly in Eulerian space, one is effectively comparing the final density with the initial one, measuring so-called propagators \cite{Crocce:2005xy,Crocce:2005xz}, which are significantly affected by the bulk flows. The EFT terms are a small correction in these statistics. Performing the comparison starting from the Lagrangian displacements we will not have to deal with this complication.

Equation (\ref{Keq}) illustrates an important difficulty in going from Lagrangian to Eulerian space. Even if one is using the effective theory approach to compute the displacement, the relation between displacement and density involves expectation values of quantities evaluated at the same point in configuration space. Arbitrarily high momenta contribute to these expectation values and thus the Lagrangian to Eulerian transformation introduces new mistakes that need to be fixed by additional counter terms. As an illustration let us consider the quadratic term in equation (\ref{Keq}): 
\be
\label{quad}
-\frac{1}{2}k_i k_j\la \Delta s_i  \Delta s_j\ra=\int_{\vec k'}\frac{(\vec k \cdot \vec k')^2}{{k'}^4}\bigl[1-\cos(\vec k' \cdot \vec r)\bigr] P(k)
\ee
Here $P(k)$  is the spectrum of the displacement divergence $\vec s_{i,i}$.
In the Zel'dovich approximation the power spectrum coincides with the linear matter power spectrum and in one loop LPT $P(k)=P_\text{lin}+2P_{13,\text{L}}+P_{22,\text{L}}$, where $2P_{13,\text{L}}$ and $P_{22,\text{L}}$ are the constituents of the one loop displacement divergence power spectra and the subscript ``L'' stands for Lagrangian.
In the Lagrangian EFT this power spectrum is corrected by a 
counterterm $\alpha k^2 P_\text{lin}$ 
due to the fact that the displacement sourced by the EFT counterterm in Eq.~\eqref{LEFTeq} correlates with the linear displacement. 
This term serves to regularize the UV-sensitivity of $P_{13,\text{L}}$. 

\begin{figure}
\includegraphics[width=0.49\textwidth]{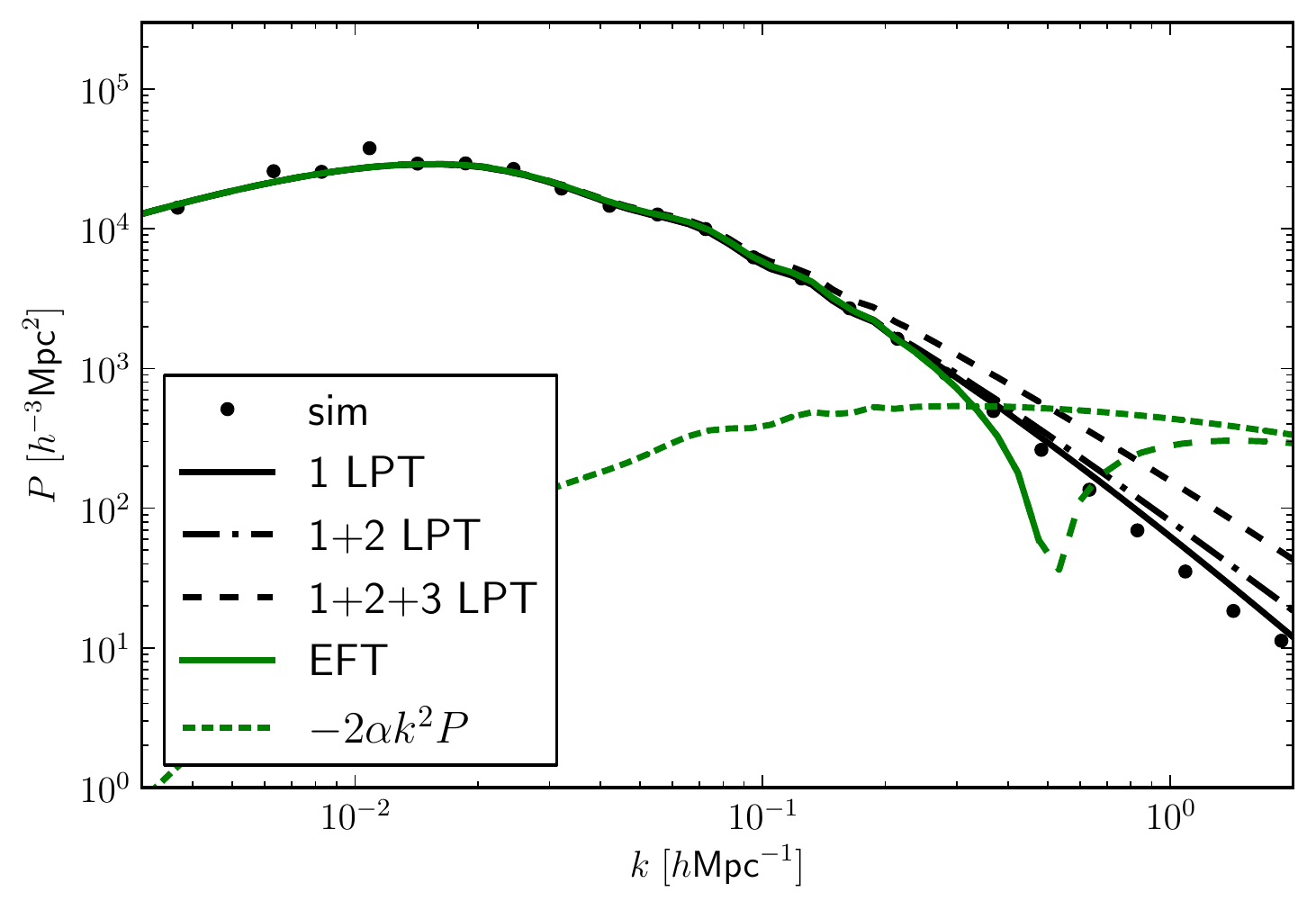}
\includegraphics[width=0.49\textwidth]{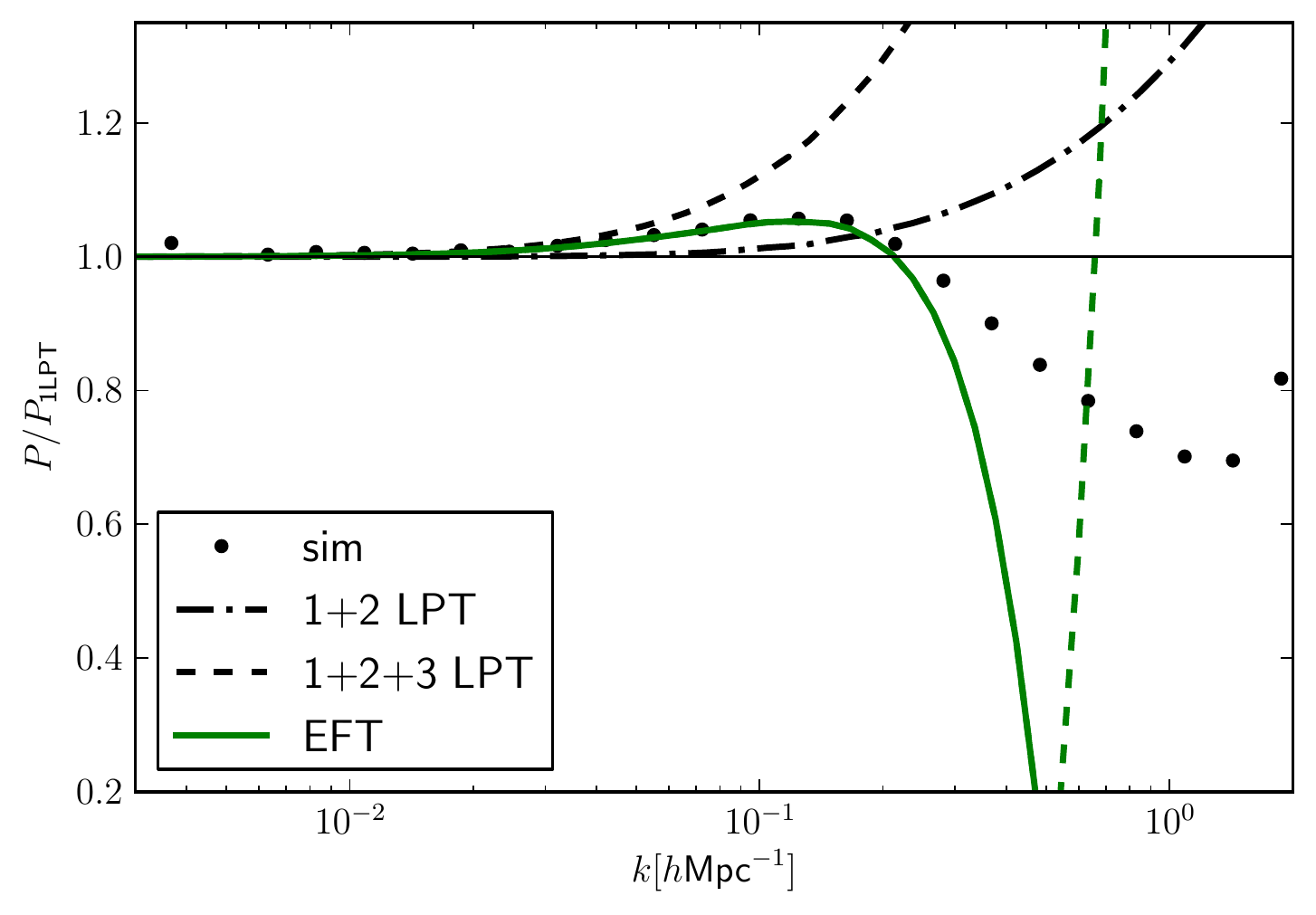}
\caption{Power spectrum of the displacement divergence $-\ii \vec k \cdot \vec s$ (left panel) and ratio of this power spectrum to linear theory (right panel). We show the non-linear displacement divergence from the simulations as well as the LPT and EFT terms. The 3LPT displacement overpredicts the true non-linear displacement. This agreement can be fixed by adding the leading order counterterm (green dotted) until $k=0.2 \ihMpc$, but for larger wavenumbers the counterterm clearly makes the agreement between simulations and theory worse.}
\label{LPTFIG}
\end{figure}

Equation (\ref{quad}) has a zero lag contributions proportional to:  
\be
\langle s_i(q) s_j(q) \rangle =  \langle s_i(0) s_j(0) \rangle = \delta_{ij}^\text{(K)} \sigma_d^2=\frac{\delta_{ij}^\text{(K)}}{3}\int_{\vec k} \frac{P(k)}{k^2}\; .
\ee
These are sensitive to the UV part of the spectrum which the EFT does not model correctly. This is illustrated in Figure \ref{LPTFIG}. The figure shows that even though on large scales the LPT displacement agrees with that computed in simulations it ceases to do so on small scales. It also shows that the range of agreement can be extended by including the EFT counter term but these new terms only improve the agreement of the displacement with simulations on large scales. On small scales they could even make the agreement of perturbative and non-linear displacements worse, which will in turn degrade the large-scale density through the zero lag contributions.

Figure \ref{fig:leftperformance} illustrates the UV sensitivity in the standard LPT calculation if one where to use equation (\ref{eq:Pk}) as is. For example if $\vec s= \vec s^{(1)}+\vec s^{(2)}+\vec s^{(3)}$ and the loop integrals are computed using a cut-off $k_\text{max}$, the result is very sensitive to that cutoff with the power spectrum differing more and more from the simulation results as $k_\text{max}$ is increased.   

The UV sensitivity generated by the mapping from Lagrangian to Eulerian arises from all terms in equation (\ref{Keq}) not just the quadratic piece we just discussed. To understand better all the different contributions it is easier to start with the expansion of the density field in terms of the displacement.  We obtain up to third order:
\be
\delta(\vec k)\approx \ii  k_i s_i(\vec k)-\frac{1}{2}k_i k_j [s_i * s_j](\vec k)-\frac{\ii}{3!}k_i k_j k_l [s_i * s_j * s_l](\vec k)\; ,
\label{eq:densexpansion}
\ee
where $[a*b]$ stands for a convolution in Fourier space.
Order by order this expression is equivalent to SPT, in particular we have at first order $\ii  k_i s_i^{(1)}(\vec k)=\delta^{(1)}(\vec k)$, at second order
\be
\delta^{(2)}=\ii  k_i s_i^{(2)}(\vec k)-\frac{1}{2}k_i k_j [s_i^{(1)} * s_j^{(1)}](\vec k)\; ,
\ee
\be
F_2(\vec p_1,p_2)=\frac{1}{2}l_2(\vec p_1,\vec p_2)+\frac{1}{2}\frac{\vec k\cdot \vec p_1}{p_1^2}\frac{\vec k\cdot \vec p_2}{p_2^2}\; ,
\ee
where $F_2$ is the second order Eulerian coupling kernel \cite{Bernardeau:2001qr}. At third order we have
\be
\delta^{(3)}=\ii  k_i s_i^{(3)}(\vec k)-k_i k_j [s_i^{(2)} * s_j^{(1)}](\vec k)-\frac{\ii}{3!}k_i k_j k_l [s_i^{(1)} * s_j^{(1)} * s_l^{(1)}](\vec k)\; ,
\label{eq:thirdorderexpansion}
\ee
\be
F_3(\vec p_1,\vec p_2,\vec  p_3)=\frac{1}{3!}l_3(\vec p_1,\vec p_2,\vec p_3)+\frac{1}{2}l_2(\vec p_1,\vec p_2)\frac{\vec k\cdot (\vec p_1+\vec p_2)}{|\vec p_1+\vec p_2|^2}\frac{\vec k\cdot \vec p_3}{p_3^2}+\frac{1}{3!}\frac{\vec k\cdot \vec p_1}{p_1^2}\frac{\vec k\cdot \vec p_2}{p_2^2}\frac{\vec k\cdot \vec p_3}{p_3^2}
\ee
However, keeping displacements up to a certain order, we automatically generate higher order terms, as we will discuss in more detail below. For the next-to-leading order power spectrum we have\footnote{For notational convenience we will often omit the momentum conserving Dirac delta $\ddir(\vec k+\vec k')$ and the normalization $(2\pi)^3$ in our equations when relating expectation values to the power spectrum. Please keep in mind that we always mean $(2\pi)^3\ddir(\vec k+\vec k')P(k)=\la \delta(\vec k)\delta(\vec k')\ra$. }
\be
P_{\delta}(k)=\underbrace{k_i k_j \la s_i | s_j\ra}_{A}+\underbrace{\ii k_i k_j k_l \la [s_i *s_j] | s_l\ra}_{B}+\underbrace{\frac{1}{4}k_i k_j k_l k_m \la [s_i *s_j] | [s_l *s_m]\ra}_{C}-\underbrace{\frac{1}{3}k_i k_j k_l k_m \la [s_i * s_j * s_l]| s_m\ra}_{D}
\ee
Let us consider the terms separately. The power spectrum of the displacement divergence $A$ receives contributions from $P_{11}$, $P_\text{13,L}$ and $P_\text{22,L}$. In LEFT, $P_\text{13,L}$ comes associated with its counterterm $\alpha k^2 P_{11}$ \cite{Porto:2013qua}.
For the bispectrum like term we have\footnote{
Here we are introducing the convolution vertex
\begin{align}
h_2(\vec p_1,\vec p_2)=&1+\vec p_1\cdot \vec p_2 \left(\frac{1}{p_1^2}+\frac{1}{p_2^2}\right)+ \frac{(\vec p_1\cdot\vec p_2)^2}{p_1^2 p_2^2}\; .
\end{align}
}
\be
B_{13}=2\ii k_ik_j k_l\la \left[ s_i^{(1)}*s_j^{(2)}\right](\vec k)\Bigr|s_l^{(1)}(-\vec k)\ra =4P(k) \int_{\vec p} \frac{1}{2}l_2(\vec k,\vec p) h_2(\vec p, \vec k-\vec p)P(p)\; .
\ee
For large internal momentum the above integral scales as
\be
-\frac{12}{35 }k^2 P(k)\frac{1}{3}\int_{\vec p} \frac{P(p)}{p^2}=-\frac{12}{35 }k^2 P(k)\sigma_d^2,
\ee
which is UV sensitive and thus needs to be regularized by another $k^2 P_{11}$ counter term beyond the one introduced in LEFT to regularize $P_\text{13,L}$. To the extent, that the form of the UV sensitivity is the same as the one in $P_\text{13,L}$, this just changes the coefficient of the $k^2P_{11}$ counterterm. As discussed above in Eq.~\eqref{eq:thirdorderexpansion} $[s^{(1)}*s^{(2)}]$ is part of $\delta^{(3)}$, and $B_{13}$ will thus be a part of $P_{13,\text{E}}$.\\
The other contribution arising from the three point correlator yields
\be
B_{22}=\ii k_i k_j k_l\la \left[s_i^{(1)}*s_j^{(1)}\right]\Bigr| s_l^{(2)}\ra =2 \int_{\vec p} \frac{1}{2}l_2(\vec p,\vec k-\vec p) h_2(\vec p,\vec k-\vec p)P(p)P(|\vec k-\vec p|)\; .
\ee
For large internal momentum the above integral scales as
\be
-\frac{2}{35} k^4\int_{\vec p} \frac{P^2(p)}{p^4}\; .
\ee
In the EFT language these UV-sensitivities that are not proportional to the linear power spectrum and scale with an external $k^4$ are called stochastic contributions \cite{Baumann:2010tm,Carrasco:2012cv}. These terms arise from a random reshuffling on small scales that conserves mass and momentum \cite{Mercolli:2013bsa,Peebles:LSS}.

Yet another stochastic contribution is given by the four point function
\be
C=k_ik_j k_lk_m\la \left[s_i^{(1)}*s_j^{(1)}\right]\Bigr|\left[ s_l^{(1)}*s_m^{(1)}\right]\ra=\frac14\int_{\vec p}h_2^2(\vec p,\vec k-\vec p)P( p)P(|\vec k-\vec p|)\; .
\ee
For large internal momentum the above integral scales as
\be
\frac{1}{20} k^4\int_{\vec p} \frac{P^2(p)}{p^4}\; .
\ee
Finally, let us discuss the cross term between the cubic and linear displacements
\be
D=\frac{1}{3}k_i k_j k_l k_m \la \left[s_i^{(1)} * s_j^{(1)} * s_l^{(1)}\right]\Bigr| s_m^{(1)}\ra=-P(k)\int_{\vec p} \left(\frac{\vec k \cdot \vec p}{p^2}\right)^2P(p)=-k^2 \sigma_{d}^2 P(k)\; .
\label{eq:dterm}
\ee
This term is proportional to the displacement variance of the field and becomes part of $P_{13,\text{E}}$. The UV sensitivity requires another counterterm of the form $k^2 P_{11}$, or equivalently, changes the coefficient of the Lagrangian counterterm.\\
The Eulerian power spectrum is given by:
\be
\begin{split}
\label{eulerP}
2P_{13,\text{E}}=&2P_\text{13,L}+B_{13}+D\\
P_{22,\text{E}}=&P_\text{22,L}+B_{22}+C
\end{split}
\ee
in particular, the low-$k$ limits of the upper line combine to the low-$k$ limit of $P_{13,\text{E}}$, which is usually regularized by the Eulerian EFT counterterm $c_\text{s}^2k^2 P$ \cite{Carrasco:2012cv}.

As we mentioned before, up to fourth order in the density field the above expressions agree with one loop SPT. However, by considering the displacement fields up to a certain order, we automatically introduce higher order corrections. One of the most straightforward one of these corrections is that in the above terms, all occurrences of $\sigma_{d,11}^2$, will be replaced by $\sigma_{d,11}^2+2\sigma_{d,13}^2+\sigma_{d,22}^2$ (with $\sigma_{d,13}^2$ being the displacement dispersion arising from $P_{13,\text{L}}$ and likewise for $\sigma_{d,22}^2$). This effectively changes the coefficient of the $k^2 P$ part of the calculation and thus modifies the coefficient of the corresponding counterterm. Besides these terms, one generates higher order fields in Eq.~\eqref{eq:densexpansion} and also higher order contributions to the other correlators discussed above. If these are large, they can not be absorbed by a $k^2P$ counterterm.

Finally, we can now return to our earlier comment on the effects of the bulk flows in the Eulerian vs. Lagrangian statistics. As an example we can consider trying to determine the EFT parameter for the one loop calculation. Our strategy in \cite{Baldauf:2015tla} for the Lagrangian displacement was to examine $P_\text{13,\text{L}}$. The  bulk flows do not contribute to $P_\text{13,\text{L}}$ but they do contribute to $P_{13,\text{E}}$ as can be seen using Eq.~\eqref{eq:dterm}. Thus if we were to apply that same strategy for the density as we did for the displacements, we would have to account for the effect of the bulk flows. This seems a bit unnecessary given that those bulk flows actually cancel. In an Eulerian calculation they cancel between  $P_{13,\text{E}}$ and $P_{22,\text{E}}$. By working with the Lagrangian displacements and keeping them exponentiated we avoid this issue altogether.  Of course there might be other approaches that one could use, but we found this strategy easy to implement especially given that we are comparing the density computed perturbatively with that computed using an $N$-body simulation, so it is very natural to work with particles and displacements, which are in close correspondence to the particles in the $N$-body simulation. 

Throughout this paper we are using a suite of $N$-body simulations discussed in BSZ. 
There are two simulation sizes: the M simulations with a box length of $500 \hMpc$ and the L simulations with a box size of $1500 \hMpc$. Besides the possibility to check simulation convergence, the different size of the simulations also imposes different Nyquist frequencies and thus different cutoffs in the perturbative displacement calculations on the simulation grid. We can use this difference to test the convergence of our results. Both simulations calculate the gravitational evolution of $1024^3$ particles from the initial redshift $z=99$ to present time. We presented numerous numerical tests of our runs in BSZ so we will not repeat them here. 

\section{One loop EFT calculation}\label{sec:oneloop}
\begin{figure}[t]
\centering
\includegraphics[width=0.49\textwidth]{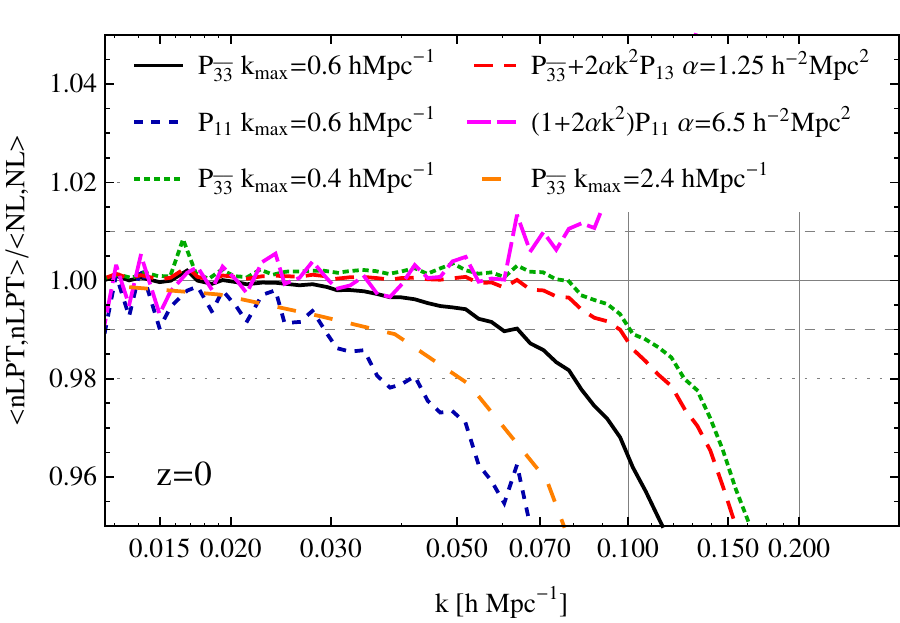}
\includegraphics[width=0.49\textwidth]{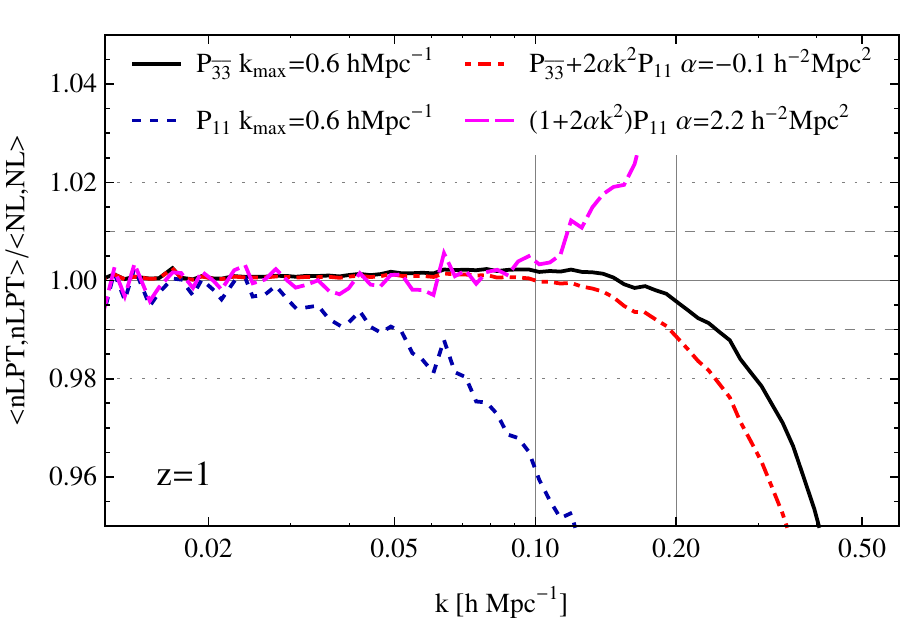}
\caption{Performance of LPT density power spectra for 1LPT/Zel'dovich and 3LPT evaluated for the same initial conditions as the simulation for redshifts $z=0$ (left panel) and $z=1$ (right panel). We clearly see the missing power in both the 1LPT and 3LPT power spectra. This failure can be reduced by setting the maximum wavenumber or LPT cutoff to $k_\text{max}=0.4\ihMpc$. The same effect can be achieved by adding a $k^2 P_{11}$ counter term as predicted in the EFT framework. The jaggedness of the 1LPT lines compared to the higher orders of LPT arises is due to the missing higher order terms that also cancel cosmic variance.}
\label{fig:leftperformance}
\end{figure}

The mapping from Lagrangian to Eulerian space is a non-linear transformation that is not additive, i.e., the effect of the $n$-th order displacements can not be calculated in isolation, but comes always associated with all lower order contributions. Throughout this paper we will thus consider density fields obtained by using the displacements computed up to a certain order of LPT and denote the highest occurring order by an overbar.  Hence, $\delta^{(\bar n)}$ is the density field calculated using all displacements up to $n$-th order. At first order there is no ambiguity and we will thus often omit the overbar for the fields arising from the Zel'dovich displacements.
A one loop calculation requires including the terms up to third order, so we start by considering\footnote{Note that the EFT parameter $\alpha$ employed here is should not be confused with the Lagrangian EFT parameter employed in BSZ.}
\be
\label{delta1L}
\delta=\delta^{(\bar 3)}+\alpha k^2 \delta^{(\bar 1)}. 
\ee
This displacement contains terms that go beyond one loop Eulerian EFT.  Because the displacement has been kept in the exponent, the expression contains high powers of $\vec s^{(1)}$, $\vec s^{(2)}$ and $\vec s^{(3)}$ that are of the same size as terms being dropped, for example from $\vec s^{(4)}$. When computing power spectra using  $\delta^{(\bar 3)}$ we also include the $\langle \vec s^{(3)}\vec s^{(3)}\rangle$ correlator that is formally a two loop term. The fact that we are keeping higher order terms is no particular problem as we are not expecting Eq.~(\ref{delta1L}) to be correct at higher order, and will only trust it in the regime where these extra terms are negligible. 

To obtain equation Eq.~(\ref{delta1L}) we have also kept only terms linear in the counter terms. This approximation should be correct in the same sense, in any case we cannot trust higher order terms arising from the exponentiation of the counter terms as higher order counter terms contribute with similar magnitude. 

One potential source of worry is that the counter terms are needed to fix  mistakes in the perturbative calculation which are kept in the exponential, while the counter terms themselves have been brought down. This would be a problem if there was a very large cancellation between the perturbative terms and the counter terms that will no longer happen after the approximation. 
 Fortunately this is not the case at one loop, the counter terms only make a small difference in that case and the inaccuracy from the expansion is not larger than the two loop terms we are neglecting at this order. 

Using  Eq.~(\ref{delta1L}) we can calculate the auto power spectrum of the model and the cross power spectrum between the model and the linear (Zel'dovich) density field,
\be
P_{\text{nl},\bar 1}=P_{\bar 1 \bar 3}+\alpha k^2 P_{\bar 1 \bar 1}\hspace{2cm}
\text{and}\hspace{2cm}
P_\text{nl}=P_{\bar 3\bar 3}+2 \alpha k^2 P_{\bar 1\bar 3}+ \alpha^2 k^4 P_{\bar 1\bar 1}\; .
\label{eq:eftmodel}
\ee
To the extend that the $\alpha^2 k^4 P_{\bar 1\bar 1}$ term is a higher order correction, we will drop it in the following. The cross correlation has the advantage, that we can explicitly probe the phases of the counterterm. This might help to avoid overfitting that could happen when using only the final power spectrum to measure the coefficients, where they could be degenerate with higher order corrections.

For reference, we will also consider the Zel'dovich density and associated power spectrum:
\be
P_{\text{nl},\bar 1}=(1+\alpha k^2) P_{\bar 1\bar 1}\; ,
\hspace{2cm}\text{and}\hspace{2cm}
P_\text{nl}=(1+\alpha k^2)^2 P_{\bar 1\bar 1}\; .
\ee
Note that we have included a counter term for Zel'dovich. As we discussed before, this is needed because even though Zel'dovich is a linear calculation, the Lagrangian to Eulerian mapping introduces zero-lag terms (for instance through Eq.~\ref{eq:dterm}) that need correction.

We show the performance of the Lagrangian EFT models in Fig.~\ref{fig:leftperformance}. We clearly see that while the bare LPT power spectra reproduce the full non-linear power spectrum on the very larges scales, they lack power on smaller scales. Despite the lack of power, the ratio of LPT and simulations is smooth, which means that IR motions have been appropriately resummed and that the BAO wiggles are well reproduced. Note in particular that the plots show the results of a single realization, i.e., by performing the LPT calculation for the same seeds we cancelled cosmic variance (see \cite{Roth:2011ru} for a similar technique employed to test SPT). This cancellation of higher order cosmic variance contributions is apparent when going from 1LPT to higher order LPT, for which the scatter is considerably reduced. This is due in particular to odd correlators (for instance $P_{12}$) that vanish only when averaging over many realizations.

\begin{figure}[t]
\includegraphics[width=0.49\textwidth]{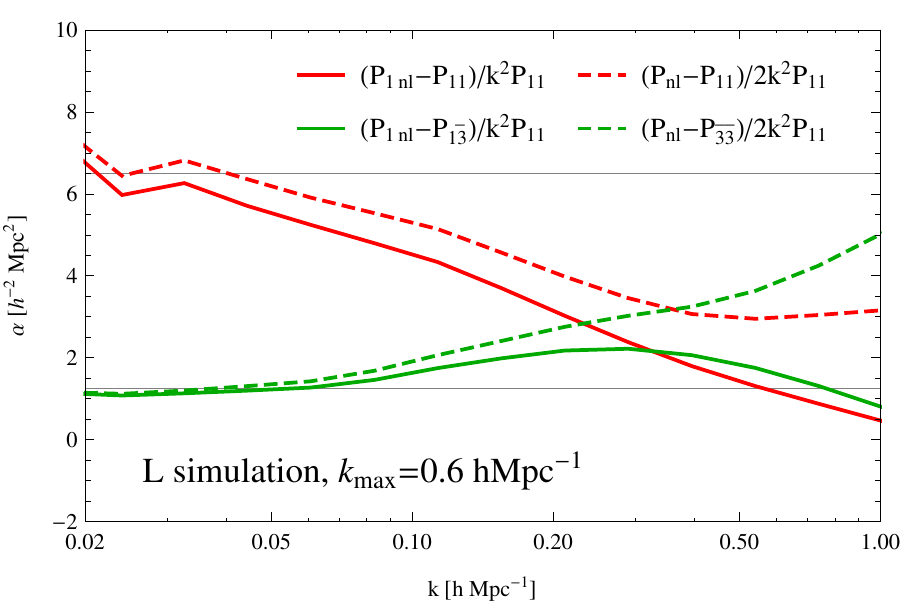}
\includegraphics[width=0.49\textwidth]{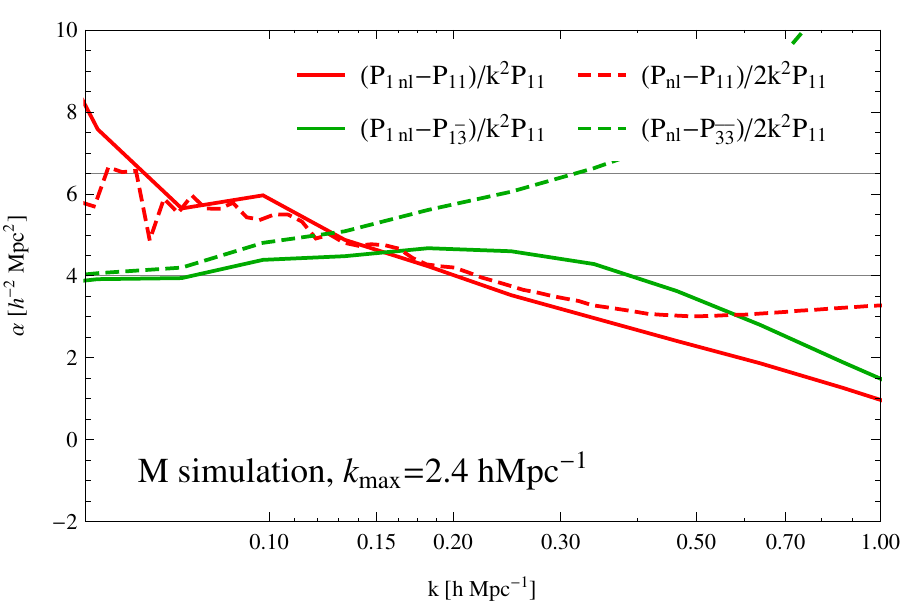}
\caption{Scale dependence of the coefficient $\alpha$ of the leading order counterterm in the L simulation (left panel) and the M simulation (right panel). We consider the one loop LPT model (green, lower) and the Zel'dovich model (red, upper). The solid lines show constraints from the cross spectrum with the Zel'dovich field, whereas dashed lines show constraints from the auto spectrum. The horizontal gray lines indicate the value of the EFT coefficient employed in Fig.~\ref{fig:leftperformance}. The difference between the green and red lines on large scales arises from the low-$k$ limit of $P_{13,\text{L}}$ ($48/63 \sigma_d^2 k^2 P$) as well as $\sigma_{d,13}^2$ and $\sigma_{d,22}^2$ entering in Eq.~\eqref{eq:dterm}. The difference between the green lines in the two panels on large scales is given by the cutoff dependence of $\sigma_{d,13}^2$ and $\sigma_{d,22}^2$.}
\label{fig:csoneloop}
\end{figure}

\begin{figure}[t]
\centering
\includegraphics[width=0.5\textwidth]{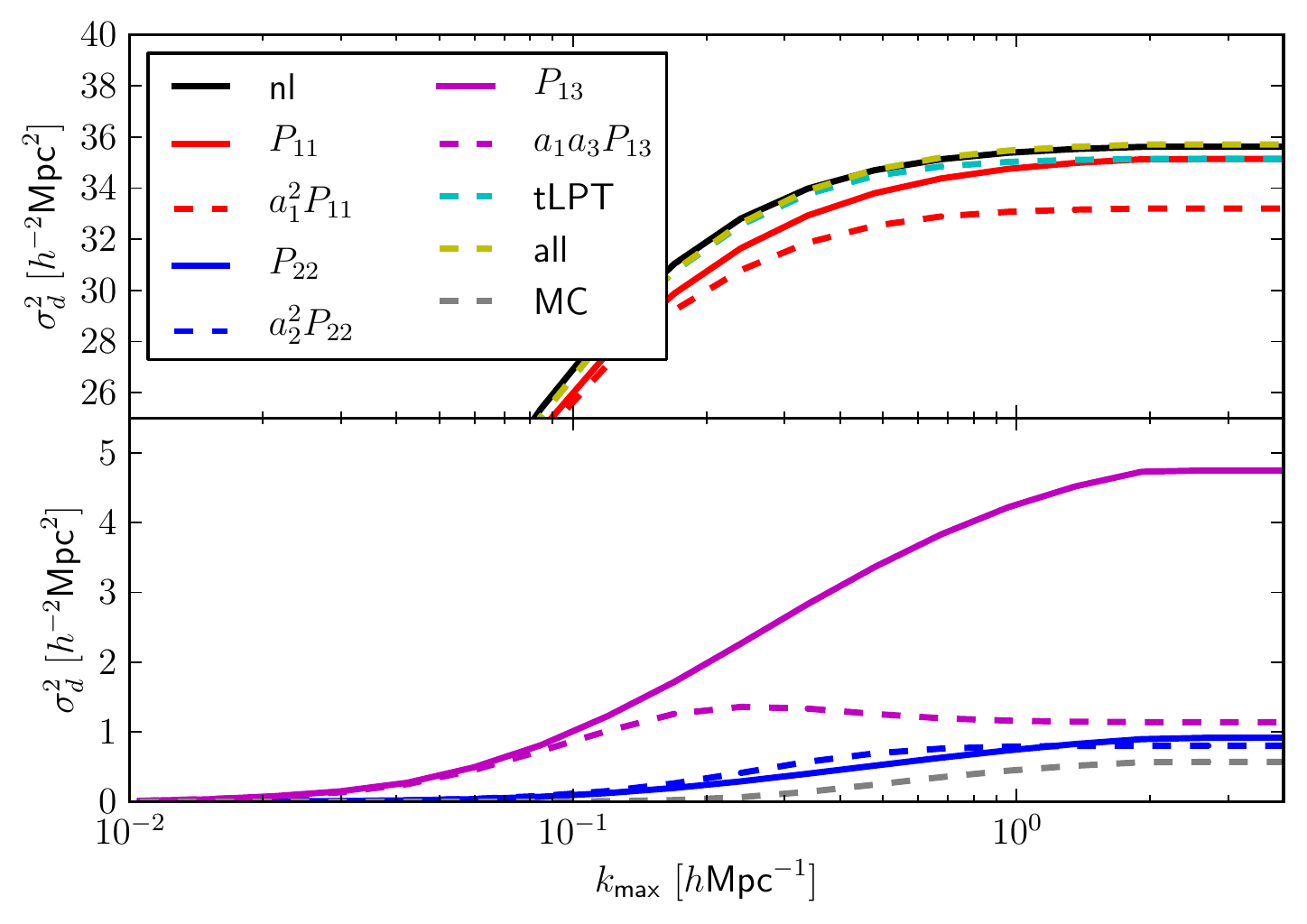}
\caption{Cumulative contributions to the displacement dispersion up to a certain maximum wavenumber $k_\text{max}$ from the various contributions to one loop LPT. We show both the bare contributions of LPT (solid) as well as the ones from the displacements regularized by the transfer functions (dashed). The non-linear displacement dispersion coincides with the linear one before a transfer function is employed. The transfer function has a mild influence on the $\sigma_{d,22}^2$ and $\sigma_{d,11}^2$ part, but significantly reduces the contribution from $\sigma_{d,13}^2$.}
\label{fig:cumulative}
\end{figure}

Fig.~\ref{fig:leftperformance} also shows that the ratios are highly cutoff dependent, especially for 3LPT with higher cutoffs leading to a stronger suppression due to the artificially large small scale displacement power in 3LPT. In this context, it is illustrative to look at the contributions to the r.m.s. displacements from modes of different scales, which are shown in Fig.~\ref{fig:cumulative}. 
In turn, by choosing an appropriately low cutoff one can bring the 3LPT prediction and the non-linear power into agreement up to $k=0.1 \ihMpc$. Let us now consider the EFT corrections to 3LPT, by choosing $\alpha$ in Eq.~\eqref{eq:eftmodel} appropriately we can improve the agreement between theory and simulations on large scales and obtain a one percent fit up to $k=0.1\ihMpc$. For comparison we also consider an EFT counter term on the Zel'dovich power spectrum only, which performs almost as well. 

The measurements of $\alpha$ as a function of scale is shown in Fig.~\ref{fig:csoneloop} for the L and M simulation. Let us first consider the counterterm on the Zel'dovich power spectrum shown by the upper pair of red lines. 
We consider the constraints from the cross-power of Zel'dovich and the non-linear density field and the constraint from the auto power spectrum of the density
\begin{align}
\alpha_\text{cross}=\frac{P_{\text{nl},\bar 1}-P_{\bar 1\bar 1}}{k^2P_{\bar 1\bar 1}}\; ,
&&
\alpha_\text{auto}=\frac{P_\text{nl}-P_{\bar 1\bar 1}}{2k^2P_{\bar 1\bar 1}}\; ,
\end{align}
which are depicted by red dashed and solid lines respectively. At the level of the displacement field they would correspond to the error and non-linear estimators employed in BSZ. We see that for both the L and M simulations the $\alpha$ parameters from both the auto and the cross estimator quickly decay from their initial amplitude of $\alpha\approx 6.5 \hMpcsq$. This scale dependence or ``running" of the EFT parameter is a first indication of the breakdown of the ansatz and the presence of higher order terms. A further indication of higher order terms playing a role is the fact that the auto and cross estimators deviate. Finally, let us comment on the size of $\alpha$ we are finding and contrast it with the typical value $\alpha^\text{E}\approx 1 \hMpcsq$ usually found when comparing the EFT prediction with the power spectrum of the density computed directly in SPT. In  Zel'dovich the linear power spectrum is damped with a factor $k^2 \sigma_d^2/2=18 k^2 \hMpcsq$, whereas in SPT one gets  $61/210 k^2 \sigma_d^2\approx 10.45 k^2 \hMpcsq$. The true damping is roughly $(61/210 \sigma_d^2+\alpha^\text{E}) k^2 \approx 11.45 k^2 \hMpcsq$ and thus one expects $\alpha=6.5 \hMpcsq$ for the Zel'dovich transfer function, which is what we see in Figure \ref{fig:csoneloop}.  This result is fairly independent of the cutoff scale since the Zel'dovich displacement dispersion converges quickly (see for instance Fig.~\ref{fig:cumulative}).

Let us now discuss the 3LPT case, for which the estimators read
\begin{align}
\alpha_\text{cross}=\frac{P_{\text{nl},\bar 1}-P_{\bar 1\bar 3}}{k^2P_{\bar 1\bar 1}}\; ,
&&
\alpha_\text{auto}=\frac{P_\text{nl}-P_{\bar 3\bar 3}}{2k^2P_{\bar 1\bar 1}}\; .
\end{align}
They are shown as the green lower set of lines in Fig.~\ref{fig:csoneloop}, and show a flattening on large scales,\footnote{The precise extraction of the EFT counterterm coefficients on large scales is complicated. While being the cleanest place to extract a low energy constant, the counterterm is only a tiny correction $\mathcal{O}(10^{-3})$ at $k=0.03\ihMpc$, and simulation codes are struggling to reproduce linear growth at the $\mathcal{O}(10^{-2})$ level \cite{Schneider:2015yka}.} which we would indeed expect over the range of validity of the ansatz in Eq.~\eqref{eq:eftmodel}. For both simulation sizes and cutoffs the constraints start to deviate from a constant and the cross and auto constraints from each other at $k\approx 0.08 \ihMpc$. This scale actually coincides with the scale where we expect two loop corrections in Eulerian Perturbation Theory to matter \cite{Baldauf:2015cs}. The difference between the low-$k$ limits of the coefficients in the L and M simulation can be explained by the strong cutoff dependence of $\sigma_{d,22}^2$ and $\sigma_{d,13}^2$.

In the right panel of Fig.~\ref{fig:leftperformance} we show the performance of the LPT and EFT terms for redshift $z=1$, for which our fiducial cutoff of $k_\text{max}=0.6\ihMpc$ leads to almost optimal results, providing a one percent accurate fit up to $k=0.2\ihMpc$ without a counterterm. As we saw from the $z=0$ example, a similar result would have been obtained with a different explicit cutoff and an appropriate counterterm. Thus there is nothing special about this particular cutoff. All we are seeing is the familiar phenomenon that the size of the counterterms runs with the cut-off and redshift. In fact the counterterms are what allows the theory to make predictions that are cut-off independent.

\section{Going beyond one loop}\label{sec:beyondone}

In BSZ we compared the displacements calculated with LPT to those produced by an $N$-body code. Although on large scales there was excellent correlation, they differed significantly on small scales. In a similar fashion as \cite{Tassev:2012cq} we defined transfer functions $a_i(k)$ such that 
\be
\vec s_\text{PT} (\vec k) = a_1(k) s^{(1)} (\vec k)+ a_2(k) s^{(2)} (\vec k)+ a_3(k) s^{(3)} (\vec k) + \cdots 
\ee
effectively minimizing the difference between the $N$-body answer and $\vec s_\text{PT} $. These transfer functions describe the motions induced by $  \vec a_\text{ct} $ in LEFT as well as higher order contributions. Their scale dependence is shown in Fig.~18 of BSZ. The full $N$-body displacements are given by 
\be
\vec s_{N-\text{body}}=\vec s_\text{PT}+ \vec s_\text{stoch},
\ee
where $\vec s_\text{stoch}$ results from $\vec a_\text{stoch}$ in LEFT.

The transfer functions on LPT displacements beyond the leading order go to zero on small scales, indicating that the LPT displacements are wrong on these scales. Without these transfer functions one-loop LPT (in particular its $P_{13,\text{L}}$ part) significantly overpredicts the r.m.s. displacements. But also the $P_{11,\text{L}}$ r.m.s. displacement is overpredicted by $1.7 \hMpcsq$, and it is a mere coincidence, that the bare value (without the appropriate displacement transfer function) agrees so well with the true non-linear r.m.s displacement. Fig.~\ref{fig:cumulative} shows the different contributions to the r.m.s. displacement. 
In  BSZ we were also able to show that with the transfer functions we include LPT contributions of orders exceeding the one explicitly computed. For example  both $P_{13,\text{L}}$ and $P_{15,\text{L}}$ contributions are captured by including the $a_1$ transfer function. We  showed that 3LPT with transfer functions (3tLPT) contains all the terms included in the two loop LPT calculation. Furthermore, we saw that fourth order LPT didn't improve the agreement with simulations because at that point the main source of error came from $\vec s_\text{stoch}$. With the transfer functions we consider in this paper we are capturing all the terms in the two-loop LPT calculation.

In this section we will consider the density fields generated from displacement fields with transfer functions. As we have argued above, the presence of additional zero lag terms in the mapping between Lagrangian and Eulerian space introduces additional mistakes. The symmetry of the EFT constrains the structure of the additional terms needed to fix those mistakes. In this section we will include these terms by introducing an additional transfer functions to the density field. The fact that the density field computed from the best possible displacements differs from the full results in ways that can be absorbed by a simple transfer function was already noted in \cite{Tassev:2012cq}. Here we extend the LPT calculation to higher orders and also make the connection with the Eulerian EFT calculations.

As in Lagrangian space, as a metric of the performance of a model for the density field (rather than the power spectrum), we consider the error power spectrum defined as 
\be
P_\text{error}=\langle |\delta_\text{nl}-\delta_\text{model}|^2 \rangle=\la\delta_\text{nl}|\delta_\text{nl} \ra+\la \delta_\text{model}|\delta_\text{model}\ra-2\la \delta_\text{nl}|\delta_\text{model}\ra\; ,
\ee
where here model refers to the perturbation theory calculation computed up to a given order. 
On a given density model we can define an overall density transfer function by
\be
T=\frac{\la \delta_\text{nl}|\delta_\text{model}\ra}{\la \delta_\text{model}|\delta_\text{model}\ra}\; ,
\ee
such that the optimal error power spectrum and the ratio of error and non-linear power spectrum are given by
\be
P_\text{error,TF}=\la\delta_\text{nl}|\delta_\text{nl} \ra-\frac{\la \delta_\text{nl}|\delta_\text{model}\ra^2}{\la \delta_\text{model}|\delta_\text{model}\ra}\; ,
\ee
\be
\frac{P_\text{error,TF}}{P_\text{nl}}=1-\frac{\la \delta_\text{nl}|\delta_\text{model}\ra^2}{\la\delta_\text{nl}|\delta_\text{nl} \ra\la \delta_\text{model}|\delta_\text{model}\ra}=1-r_\text{cc}^2\; ,
\ee
where $r_\text{cc}$ is the cross correlation coefficient.
This statistic measures the r.m.s. deviations between a certain model and the non-linear data. It is quantifying the performance of the model at the field level rather than the level of the power spectrum.

\begin{figure}
\includegraphics[width=0.49\textwidth]{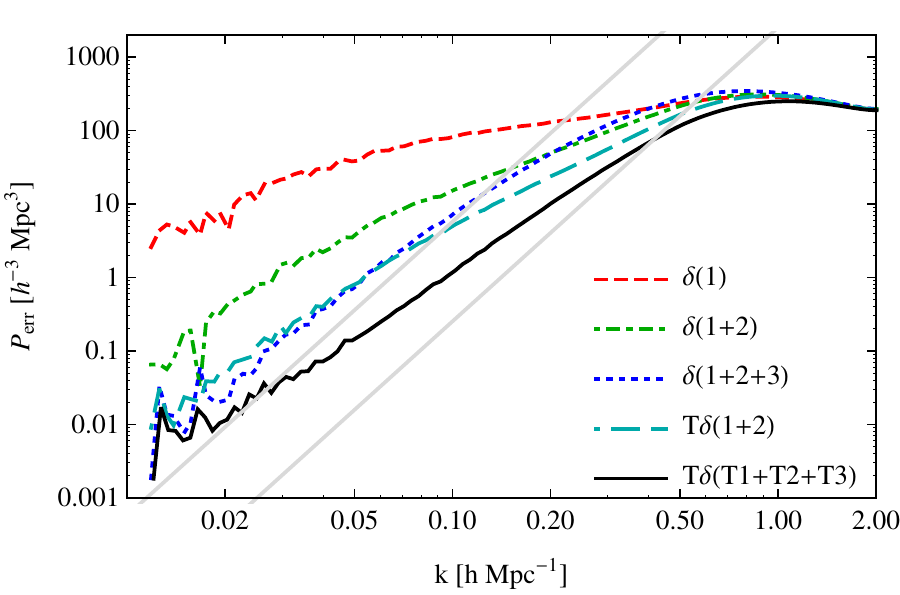}
\includegraphics[width=0.49\textwidth]{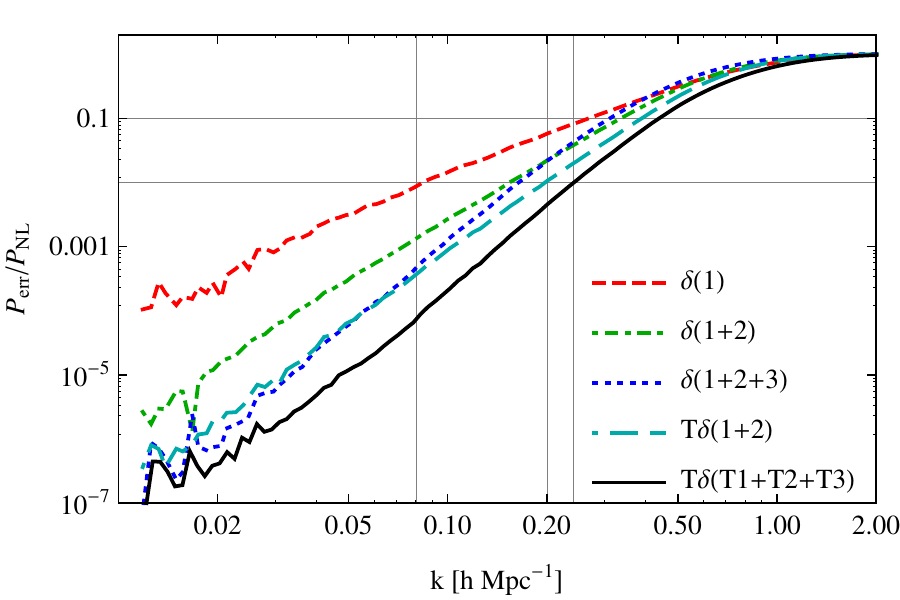}
\caption{Error power spectra at redshift $z=0$ for the L simulation and various orders of LPT and tLPT. \emph{Left panel: }Total power. The gray lines show the expectation for the stochastic term for $k_\text{nl}=0.3\ihMpc$ and $k_\text{nl}=0.5\ihMpc$, respectively. We see a decrease in mode coupling as we go to higher orders and implement transfer functions on the density. \emph{Right panel: }Ratio of the error power spectrum and the full non-linear power spectrum.}
\label{fig:errorTF}
\end{figure}

Fig.~\ref{fig:errorTF} shows the error power spectrum for different perturbation theory calculations. As an illustration we show the results of LPT without any transfer functions (EFT counterterms). One striking observation in this case is that including $\vec s^{(3)}$ makes things worse. The curve labeled $T\delta(1+2)$ is effectively the equivalent of the LEFT one loop calculation. Even if $\vec s^{(3)}$ is not included, the only part of $\vec s^{(3)}$ relevant at this order is the cross term $1-3$ which is captured by the transfer function. We have also computed another version of an effective one loop calculation $T\delta(1+2+3)$ which gives very similar results (although the transfer function is different). The comparison between $T\delta(1+2)$ and $\delta(1+2+3)$ illustrates the fact that the overall transfer function is indeed crucial, it is better to include that than to add an additional order in the displacement field if the goal is to minimize the error at the level of the density field. 

The line labeled $T\delta(T1+ T2+ T3)$ uses transfer functions for all the displacements up to the third order and plus an overall transfer function on the density to correct the leading order part of the zero-lag terms from the mapping.  This model yields the lowest error power spectrum, such that we interpret it as the best possible perturbative model.
We have also compared this to the case where we have several transfer functions at the level of the density determined in the same way as the displacement transfer functions in BSZ: $T_3\delta(T1+ T2+ T3) + T_2\delta(T1+ T2)+ T_1\delta(T1)$ (see App.~\ref{sec:crosscheck}). The latter contains all the terms in a two-loop EFT calculation and gives very similar results to the cases considered here.\footnote{At the two loop level, the linear counter term enters correlated with itself as $P_{\tilde1 \tilde 1}$, but the second and third order counterterms enter only correlated with perturbative terms as $P_{\tilde2 2}$ and $P_{\tilde3 1}$. To this extent, the counter term if present in the data, should correlate with the perturbative second and third order basis vectors. Thus our $T_ 3\delta(T1+ T2+ T3) + T_2\delta(T1+ T2)+ T_1\delta(T1)$ captures all relevant terms for two loops. In fact one expects this example to be if anything better than one might do in a first principle Eulerian EFT calculation as we are allowing the transfer functions to have an arbitrary shape. We will discuss this in more detail in App.~\ref{app:equiv}.} The error we see in $T\delta(T1+ T2+ T3)$ does not decrease appreciably if we include higher order displacements either. We interpret this error as arising primarily from the stochastic displacements $\vec s_\text{stoch}$. 

We had uncovered a stochastic contribution to the displacement already in BSZ. The stochastic term in the EFT arises from a mass and momentum conserving shuffling of mass on small scales. As a result, its power spectrum has to scale as $k^4$ for low wavenumbers and we saw this behavior clearly in BSZ for the  divergence of the displacement. Fig.~\ref{fig:errorTF} however indicates that this behavior seems to be violated by the error power spectrum in our simulations even on rather large scales, which we are identifying with the stochastic term. We have performed a number of numerical tests to check the stability of this stochastic contribution and found it to be stable. We show some of the checks in the Appendix, for example Fig.~\ref{fig:errorLvsM} shows the comparison of the error power spectra found when analyzing the L and M simulations.

We will discuss the shape of the error power spectrum in more detail in the next section. What we will see is that  on very large scales, this stochastic term of the density indeed agrees with the power spectrum of the stochastic displacement divergence, as it should. But we will also identify corrections that arise from the mapping from Lagrangian to Eulerian space leading to deviations from the $k^4$ scaling on surprisingly large scales ($k\approx 0.03\ihMpc$ at $z=0$).

\begin{figure}[t]
\centering
\includegraphics[width=0.49\textwidth]{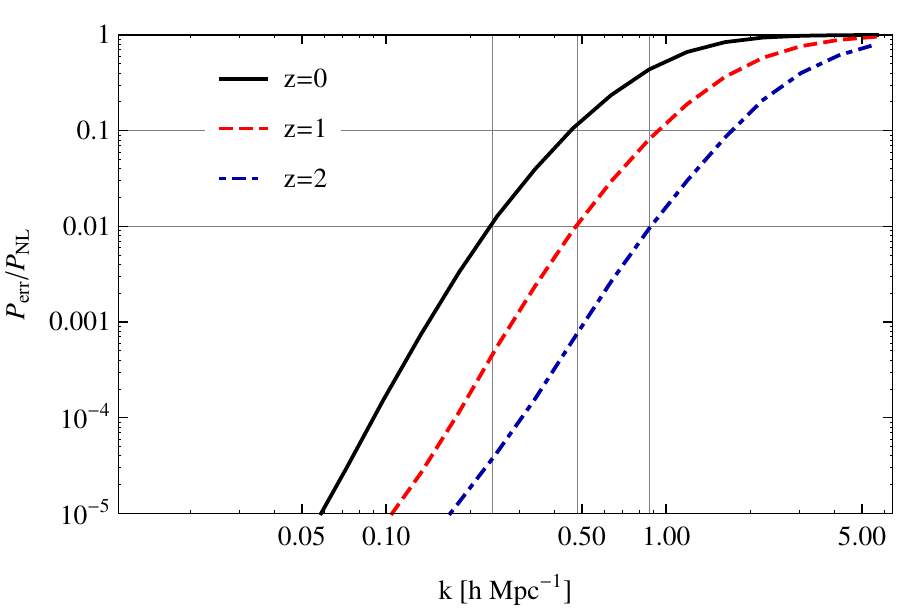}
\includegraphics[width=0.49\textwidth]{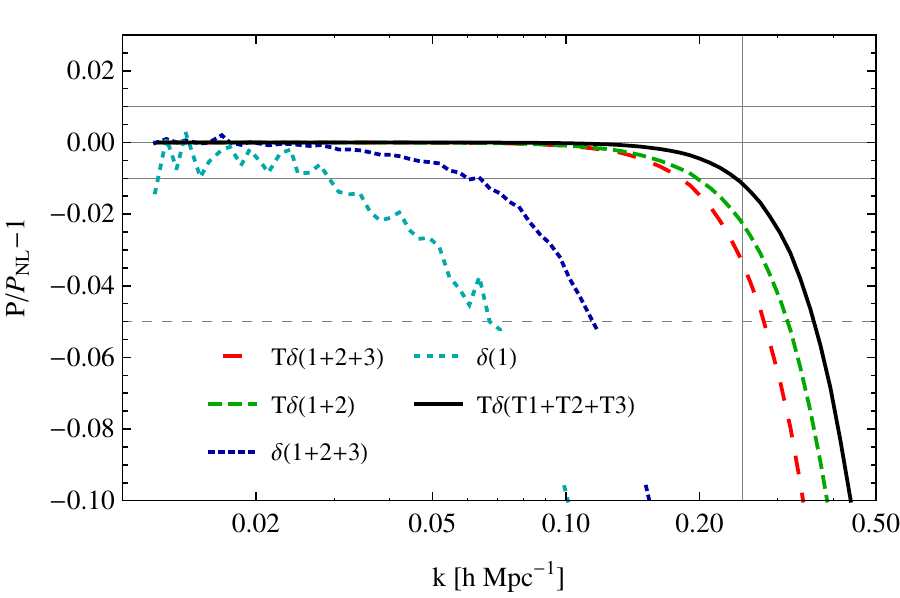}
\caption{\emph{Left panel: }Ratio of the best possible EFT power spectrum to the non-linear power spectrum as a function of redshift. We indicate the $1\%$ and $10\%$ accuracy lines and mark the crossing of the $1\%$-threshold by vertical lines, whose wavenumbers are given in Tab.~\ref{tab:percentlevels}. \emph{Right panel: }Ratio of the perturbative model with and without transfer functions and the non-linear power spectrum at $z=0$.}
\label{fig:reach}
\end{figure}

\begin{table}[b]
\centering
\begin{tabular}{r|r|r}
$z$ & $k_{1\%}$ & $k_{10\%}$\\
\hline
0 & $0.25\ihMpc$& $0.46 \ihMpc$\\
1 & $0.48\ihMpc$ & $0.98\ihMpc$\\
2 & $0.85\ihMpc$& $1.72 \ihMpc$
\end{tabular}
\caption{Wavenumbers, where the stochastic term amounts to a $1\%$ or $10\%$ correction to the non-linear matter power spectrum.}
\label{tab:percentlevels}
\end{table}

In Fig.~\ref{fig:reach} we show the ratio of error and non-linear power spectrum $P_\text{err}/P_\text{NL}$ for three redshifts $z=0,1,2$ for the $T\delta(T1+ T2+ T3)$ example to quantify up to which wavenumber the perturbative calculation can be expected to agree with the $N$-body result. We quote the wavenumbers at which the stochastic power crosses the $1\%$ and $10\%$ level in Tab.~\ref{tab:percentlevels}. While one should not focus too much on the specific values, one should definitely note the steepness of the curves in the left panel of Fig.~\ref{fig:reach}. This means that at a fixed $k$ away from the non-linear scale, the size of the error changes dramatically as one goes to higher wavenumbers. This is important, since for data analysis applications, such as trying to see the small effects of primordial non-Gaussianity in the two- and three point functions \cite{Alvarez:2014vva}, precision will probably be more important than reach. 

The right panel of Fig.~\ref{fig:reach} shows $P_\text{model}/P_\text{NL}-1$ and illustrates again that the biggest improvement in reach comes from the inclusion of the final transfer function, fixing the problems caused by the mapping. The comparison between $T\delta(1+2)$ and $T\delta(1+ 2+3)$, which are both effectively equivalent to one-loop EFT calculations (with higher derivative counterterms) shows the difference that the higher order terms that are only partially included can make. It is amusing to note that $T\delta(1+ 2+3)$ is actually slightly worse, so the additional work to include $\vec s^{(3)}$ did not result in an improvement here. This is perhaps not surprising given how bad $\vec s^{(3)}$ is on small scales and the fact that the Lagrangian to Eulerian mapping makes the large scale density depend on these mistakes. Of course with additional freedom from more counter terms one should be able to absorb these differences. Finally one may notice that in terms of reach, 
$T\delta(T1+ T2+ T3)$ does not even improve by a factor of two. But reach is perhaps the wrong metric as the error curves are very steep. Fig.~\ref{fig:errorTF} shows that away from the non-linear scale, the error in 
$T\delta(T1+ T2+ T3)$ is smaller than the one in $T\delta(1+ 2)$ by about one order of magnitude. 

\begin{figure}[t]
\centering
\includegraphics[width=0.89\textwidth]{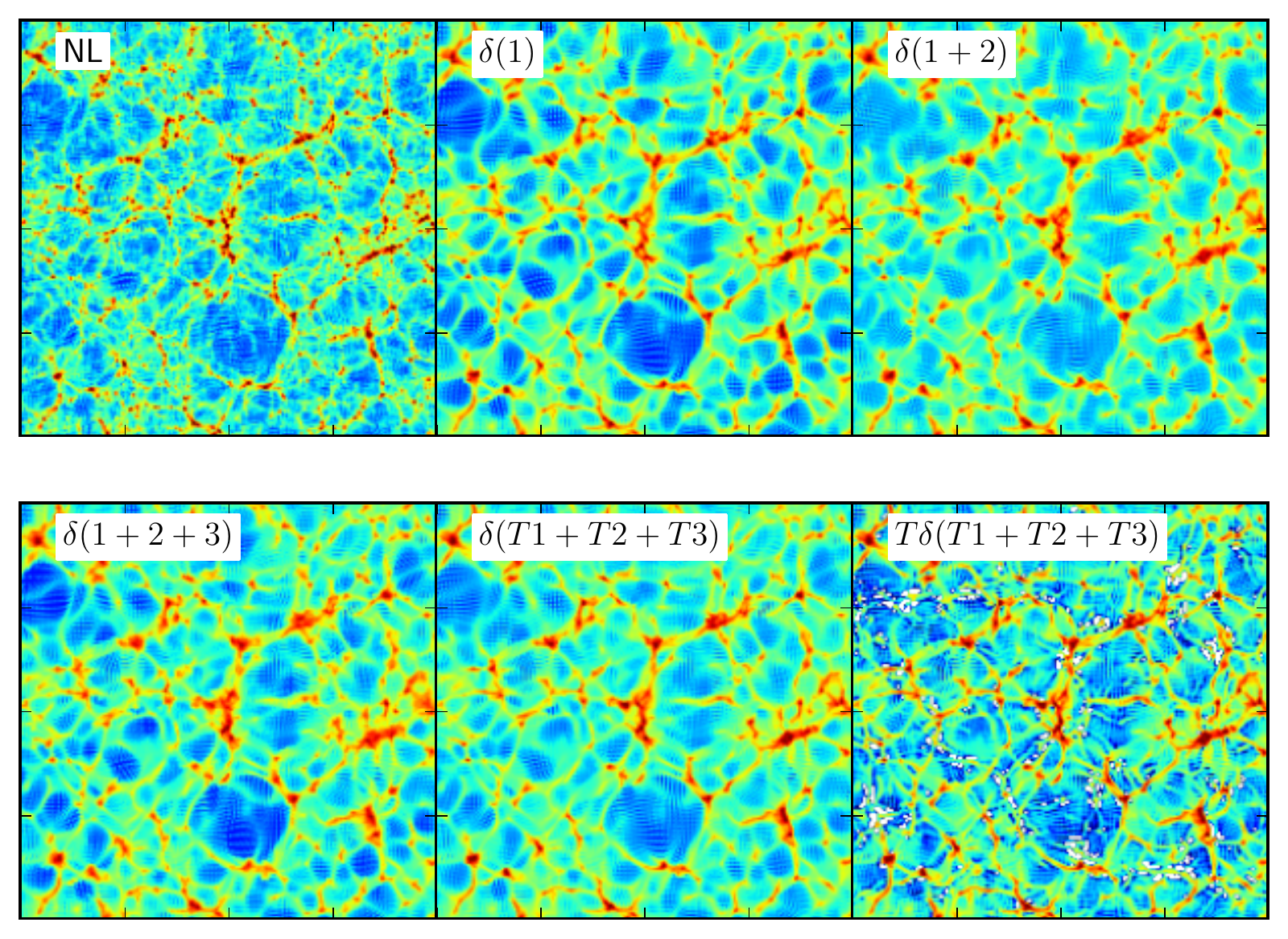}
\caption{Non linear transformation of the density field in a patch of $300\hMpc$ length and $15 \hMpc$ depth.}
\label{fig:densities}
\end{figure}

A map of the various density fields discussed in this section is shown in Fig.~\ref{fig:densities}. It clearly shows how well correlated the structure in a Zel'dovich realization is with the non linear structure. Overdensities are washed out and voids are clearly underdense. Adding higher order displacement fields and transfer functions on the displacement fields has no strong imprint in this picture beyond a slight sharpening of the overdensities and filaments. The final density transfer function shown in the last panel clearly has the strongest effect, most remarkably a sharpening of structures in all environments. But even at this level there are still obvious differences between the best perturbative approach and the non-linear field.

\section{Stochastic Term}\label{sec:stoch}
In BSZ we identified an irreducible error at the field level that we associated with the stochastic term of the EFT. In this Section, we are relating the Lagrangian stochastic term to the Eulerian one. From now on we will consider displacement fields up to a certain order including all transfer functions and denote them $\vec s_\text{PT}$, in particular we will be mostly concerned with the displacement fields up to third order, {\it i.e.}, $\vec s_\text{PT}=a_1 \vec s^{(1)}+a_2 \vec s^{(2)}+a_3 \vec s^{(3)}$. The total displacement field is then the sum of the perturbative and the stochastic part $\vec s=\vec s_\text{PT}+\vec s_\text{stoch}$.

\subsection*{Toy model: origin of the transfer function for the density}
Let us consider the case where we expand only the stochastic displacement in Eq.~\eqref{eq:densmap}
\be
\begin{split}
(2\pi)^3\ddir(\vec k)+\delta(\vk)\approx \int \derd^3q \eh{\ii \vk \cdot (\vec q+\vec s_\text{PT})}\Biggl(&1+\ii k_i s_{\text{stoch},i}-\frac12  k_i k_j s_{\text{stoch},i}s_{\text{stoch},j}\\
& -\frac{\ii}{3!}   k_i k_j k_l s_{\text{stoch},i}s_{\text{stoch},j}s_{\text{stoch},l}\Biggr)
\end{split}
\ee
The stochastic term is likely related to small scale non-linear phenomena and is orthogonal to the perturbative terms by construction. Averaging the above equation over the short modes and assuming that there are no correlations between the statistical properties of the stochastic term and the long modes we get:
\begin{align}
(2\pi)^3\ddir(\vec k)+\la\delta(\vk)\ra_\text{stoch}&\approx \int \derd^3q \eh{\ii \vk \cdot (\vec q+\vec s_\text{PT}(\vec q))}\left(1-\frac12  \la(\vk\cdot\vec s_\text{stoch}(\vec q))^2\ra_\text{stoch} \right)
\end{align}
This result could have been obtained at the level of the exponential using the cummulant expansion theorem and neglecting coupling between short and long modes
\be
(2\pi)^3\ddir(\vec k)+\la\delta(\vk)\ra_\text{stoch}\approx \int \derd^3q \eh{\ii \vk \cdot (\vec q+\vec s_\text{PT})}\eh{-\frac{1}{2}k^2 \sigma_{d,\text{stoch}}^2}
\ee
and thus
\be
\delta(\vec k)=\delta_\text{PT}(\vec k)\eh{-\frac{1}{2}k^2 \sigma_{d,\text{stoch}}^2}
\ee
This is a Gaussian smoothing of the PT results, it has the form of a quadratic counterterm and its amplitude is given by the stochastic displacement $\sigma_{d,\text{stoch}}^2 \approx 0.7\hMpc$. It also indicates, that for the best density field one should consider the best possible result for the displacement field and then multiply with an overall transfer function that captures the r.m.s. displacement of the stochastic term.
Of course there are additional contributions when one takes into account that the statistical properties of the stochastic term depend on  the long modes. At the lowest order this was discussed already in \cite{Porto:2013qua} and found empirically in \cite{Tassev:2012cq}, here our goal was to illustrate this point again for convenience of the reader. 

\subsection*{Mapping of the stochastic terms}
Let us reconsider the mapping between Lagrangian displacements and Eulerian density Eq.~\eqref{eq:densmap}, distinguishing between PT and stochastic terms
\be
\begin{split}
(2\pi)^3\ddir(\vec k)+\delta(\vk)\approx \int \derd^3q \eh{\ii \vk \cdot \vec q}\Biggl(&1+\ii k_i (s_\text{PT}+s_\text{stoch})_i-\frac12  k_i k_j (s_\text{PT}+s_\text{stoch})^2_{ij}\\
&-\frac{1}{3!} \ii  k_i k_j k_l(s_\text{PT}+s_\text{stoch})^3_{ijl} \Biggr)\; .
\end{split}
\ee
The relevant contributions correcting the Lagrangian stochastic term are then given by
\be
\begin{split}
P_\text{stoch,E}(k)=&k_i k_j \la s_{\text{stoch},i}|s_{\text{stoch},j} \ra-2 \ii k_i k_j k_l \la [s_{\text{stoch},i} * s_{\text{PT},j}] | s_{\text{stoch},l} \ra-\ii k_i k_j k_l \la [s_{\text{stoch},i} * s_{\text{stoch},j}] | s_{\text{stoch},l} \ra\\
+&k_i k_j k_l k_m \la [s_{\text{stoch},i} * s_{\text{PT},j}] | [s_{\text{stoch},l}*s_{\text{PT},m}] \ra+k_i k_j k_l k_m \la [s_{\text{PT},i} * s_{\text{PT},j}*s_{\text{stoch},l}] | s_{\text{stoch},m} \ra
\end{split}
\ee
Using the fact that the mode coupling is orthogonal to perturbation theory, we can readily calculate the contributions from the even correlators in the above calculation. In particular, we have for the first, fourth and fifth term
\be
\begin{split}
P_\text{stoch,E}(k)\supset& P_\text{stoch,L}(k)+P_\text{stoch,L}(k)\int_{\vec p} \frac{(\vec k\cdot \vec p)^2}{p^4} P_{11}(p)+\int_{\vec p}  \left[ \frac{\vec k \cdot\vec p}{p^2} \frac{\vec k \cdot(\vec k-\vec p)}{(\vec k-\vec p)^2}\right]^2  P_\text{stoch,L}(p) P_{11}(\vec k-\vec p)\; ,\\
=&P_\text{stoch,L}(k)\bigl(1-\sigma_{d,11}^2 k^2\bigr)+\int_{\vec p}  h_2(\vec p, \vec k-\vec p)  P_\text{stoch,L}(p) P_{11}(\vec k-\vec p),
\end{split}
\ee
where we assumed Gaussianity for the stochastic term. Unfortunatly, this calculable correction is too small to explain the differences between the two stochastic terms identified in the simulations in Lagrangian and Eulerian space.

In BSZ we were only concerned with the two point functions involving the fields and not the three point functions. Thus, we only established the orthogonality of stochastic term and perturbative terms at the level of the two point function and did not consider the three point function of the stochastic field itself or possible three point functions between the stochastic and the perturbative fields.
These could arise if for example, the stochastic piece is modulated by the perturbative modes, leading to a coupling\footnote{For illustrative purposes we will use the shorthand notation $\delta=-\ii \vec k\cdot \vec s$.}
\be
\delta_\text{stoch}^{(2)}(\vec k)=\int_{\vec p}K_2(\vec p,\vec k-\vec p)\delta^{(1)}_\text{stoch}(\vec p)\delta_\text{PT}(\vec k-\vec p)\; .
\ee
with an unknown coupling kernel $K_2$. This term does not correlate with pure perturbation theory terms and does thus not violate our requirement that $\la\delta_\text{PT}|\delta_\text{stoch} \ra=0$.
Alternatively there can be a correlation of the stochastic term with the square of the stochastic term, that can be encoded by another kernel and the replacement $\delta_\text{PT}\to \delta_\text{stoch}$ in the above equation. This is indeed quite natural for a non-Gaussian field generated by non-linearities.
With the above couplings we would have for the three point functions
\be
\la \delta_\text{PT}\delta_\text{stoch}|\delta_\text{stoch}\ra=\int_{\vec p}K_2(\vec p,\vec k-\vec p)P_\text{stoch}(\vec p)P_{11}(\vec k-\vec p)
+P_\text{stoch}(\vec k)\int_{\vec p}K_2(\vec k,\vec p)P_{11}(p)
\ee
 We have no a priory knowledge about the coupling kernels or the three point functions of the stochastic term and thus have to extract the three point function from simulations.

\begin{figure}[t]
\centering
\includegraphics[width=0.49\textwidth]{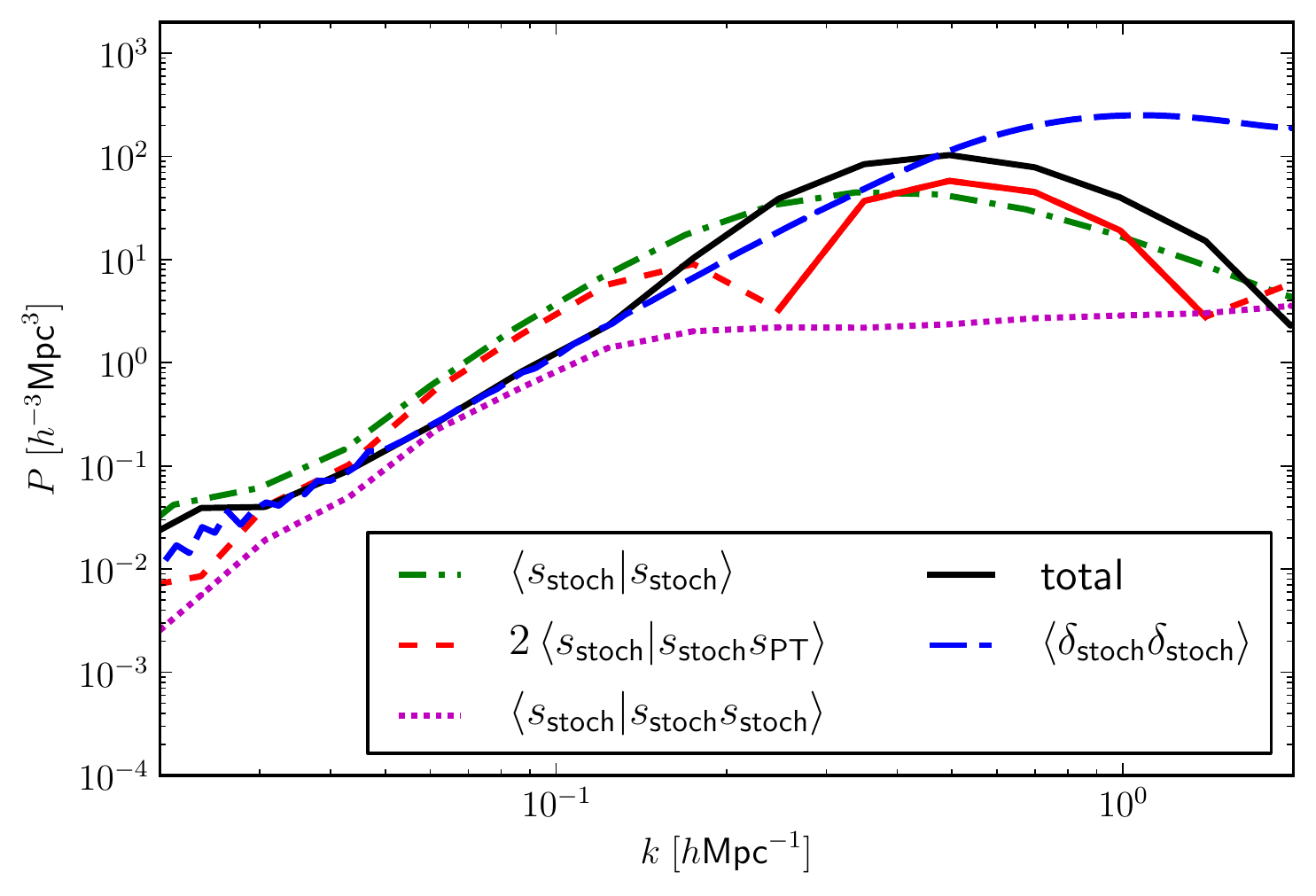}
\caption{Contributions to the mapping from Lagrangian to Eulerian space. We start from the Lagrangian mode coupling (green dash-dotted), to which we add the next to leading corrections $\la s_\text{stoch}|[s_\text{stoch}*s_\text{PT}]\ra$ and $\la s_\text{stoch}|[s_\text{stoch}*s_\text{stoch}]\ra$. There is a significant cancellation between these two terms themselves and with the original mode coupling term. The sum of these three terms (black) traces the Eulerian mode coupling defined in the previous section (red dot-dashed) very well on large scales.}
\label{fig:mappingterms}
\end{figure}

We measured the correlators $\la \delta_\text{PT}\delta_\text{stoch}|\delta_\text{stoch}\ra$ and $\la \delta_\text{stock}\delta_\text{stoch}|\delta_\text{stoch}\ra$ for the stochastic term identified in our simulations and show the results in Fig.~\ref{fig:mappingterms}. We clearly see that the Lagrangian stochastic term for the displacement dispersion differs from the Eulerian one on all but the largest scales. In particular, the Eulerian stochastic term is smaller than the Lagrangian one for $k<0.25 \ihMpc$ and exceeds it for higher wavenumbers. This behaviour is qualitatively expected from the mapping between Lagrangian and Eulerian coordinates and the collapse of structure. More quantitatively, we consider the corrections from the correlator of the stochastic field with the product of perturbative and stochastic fields at a different location and the correlator of the stochastic field and the square of the stochastic field at another position. The first of these two terms, leads to order unity negative corrections to the original stochastic power around $k\approx 0.1\ihMpc$, such that only the bispectrum of the stochastic term survives. The sum of the original term and the two leading corrections yields a result that agrees with the stochastic term identified in the previous section, by projecting out the perturbative part of the density field. This is on the one hand very reassuring, since we have recovered the stochastic part of the density field in two independent ways. On the other hand, the importance of three point functions in the mapping from Eulerian and Lagrangian makes it very hard to assess the size of the Eulerian stochastic term based on the Lagrangian two point function only. 

The most important observation from the EFT point of view is that the stochastic term asymptotes to its expected leading $k^4$ scaling only on very large scales $k\approx 0.01\ihMpc$ and has a shallower slope where it matters at the percent level $k\approx 0.2\ihMpc$. This complicates its modelling in the EFT framework, since higher orders in derivatives or additional scales besides the non-linear scale need to be considered.

\subsection*{The stochastic term in the halo model perspective}
The halo model (for a review see \cite{Cooray:2002dia}) splits the mass distribution in the Universe into distinct subsets or regions, each with its center of mass and a density profile. The correlations are then disentangled into correlations between these objects and correlations within the objects. Since the objects are often taken to be dark matter haloes, these terms are known as one- and two-halo terms, respectively. Clearly, the correlations within haloes are non-perturbative and the corresponding halo profiles are thus usually extracted from numerical simulations or fitting functions to the latter. The only perturbative part is the correlation between the centers of the overdense regions. In perturbation theory, there is no such distinction. On the contrary, whenever one employs models for the clustering of the dark matter field, one predicts the field at all positions, even within highly overdense regions, such as dark matter haloes. Thus the perturbation theory result contains a one halo term, that is clearly wrong, since there is no hope that a perturbative approach will recover the virialized structure. This was previously discussed in detail in \cite{Valageas:2010yw}. Here we would like to discuss the sizes of the terms in the context of our measurements.

The power spectrum of terms within a single halo is given by
\be
P_\text{1H}(k)=\int \derd M n(M) \frac{M^2}{\bar \rho^2} u^2(k)\; ,
\label{eq:onehalterm}
\ee
where $n(M)$ is the halo mass function and $u$ is the normalized profile satisfying $u(k)\xrightarrow{k\to 0}1$.
The amplitude of the one halo term is thus given by the mean squared density fluctuations
\be
A_{1\text{H}}=\int \derd M n(m) \frac{M^2}{\bar \rho^2}\approx 330 h^{-3}\text{Mpc}^3\; .
\ee
The above integral is dominated by massive dark matter haloes $M\approx 10^{14}\; h^{-1}M_\odot$.

Assuming that perturbation theory correctly predicts the center of mass of haloes, the remaining profile error between the perturbative and non-linear profiles $u_\text{PT}$ and $u_\text{NL}$ will contribute the following mistake
\be
\Delta P_\text{1H}=\int \derd M\; n(M) \frac{M^2}{\bar \rho^2} \Bigl[u_\text{NL}(k)-u_\text{PT}(k)\Bigr]^2\; .
\label{eq:profilediff}
\ee
Imposing mass and momentum conservation on $u_\text{NL}-u_\text{PT}$, the $k^0$ and $k^1$ components cancel from this term and it thus starts as $k^2$. Thus we have finally that $\Delta P_\text{1H}\propto k^4$ as one should expect for a mass and momentum conserving local process. This does not mean that the one halo term by itself is compensated, just the difference between the PT and non-linear profiles is compensated. 

\begin{figure}[t]
\centering
\includegraphics[width=0.49\textwidth]{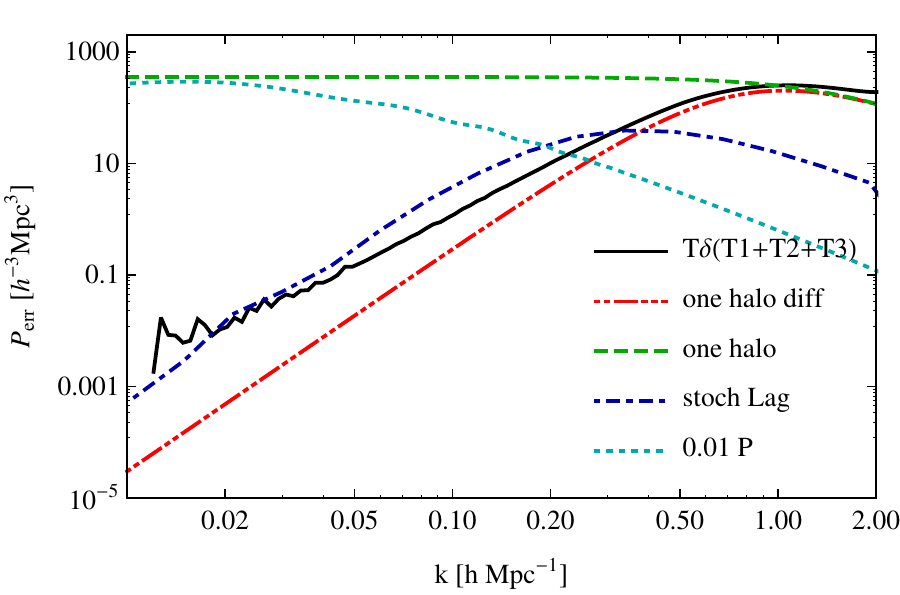}
\caption{Power spectrum of the non-linear one halo term Eq.~\eqref{eq:onehalterm} extracted from the simulations and the one halo term of the profile difference between the simulation and 1LPT Eq.~\eqref{eq:profilediff}. The gray line shows $1\%$ of the linear power spectrum. We can clearly see that the profile differences amount to percent level corrections to the linear power spectrum at $k\approx0.3\ihMpc$. We also show the stochastic term in Lagrangian space and Eulerian space. The Eulerian stochastic term is larger than the profile difference on large scales but approaches it at $k\approx 1 \ihMpc$.}
\label{fig:haloprof}
\end{figure}

We identified haloes in our L simulation using a Friends-of-Friends halo finder with linking length 0.2. We then extracted the non-linear halo profiles and the profiles of the particles displaced by Zel'dovich rather than non-linear dynamics. Phenomenologically, the halo constituent particles are roughly at the right position but the PT halo is more  dispersed.
Fig.~\ref{fig:haloprof} shows the profile differences between the true non-linear profiles and the Zel'dovich profiles as well as the non-linear profile itself. Above $k\approx 1\ihMpc$ the two agree, but on larger scales the compensation kicks in and the the profile difference asymptotes to the expected $k^4$ behaviour. If modeled according to the self similar scaling $\Delta_\text{stoch}=(k/k_\text{nl})^7$ \cite{Pajer:2013jj}, this term corresponds to $k_\text{nl}=0.5\ihMpc$. 
This profile difference is clearly a lower limit on the size of the stochastic density power spectrum since other non-linear structures (walls, filaments and voids) and the stochastic error on the center of mass will contribute as well. Given the large scales/small wavenumbers where the stochastic term asymptotes to $k^4$, it is clearly not dominated by dark matter halo profiles at these scales. Yet, the scale where the halo profile difference amounts to one percent of the linear power spectrum is $k=0.28 \ihMpc$, which is close to the scale where the stochastic term crosses the $1\%$ threshold.
Recently, \cite{Seljak:2015rea} found a compensated one halo term that is orders of magnitude larger than the term discussed here. This difference arises from the fact that we are taking the difference between 1LPT and the simulations at the field level, whereas it was taken at the level of power spectra in \cite{Seljak:2015rea}.

\section{Conclusions}\label{sec:concl}
Using the IR resummed basis provided by Lagrangian Perturbation Theory, we compared the perturbative density fields with the results of $N$-body simulations sharing the same initial conditions. 

Using LPT generated displacement fields with a $k^2$ transfer function on the linear field, as suggested by the EFT at leading order, we manage to model the non-linear power spectrum to $1\%$ precision up to $k\approx 0.1 \ihMpc$ at $z=0$. Precise measurements of the EFT coefficient show a scale dependence of the coefficient extracted at 3LPT level for $k>0.07\ihMpc$ that is likely due to the presence of missing two loop corrections. We also showed that the leading EFT counterterm is able to capture the cutoff dependence of the LPT density field for the two cutoffs $k_\text{max}=0.6\ihMpc$ and $k_\text{max}=2.4 \ihMpc$ considered here.
To assess the maximum range of validity of perturbative approaches for the density field and in order to avoid issues with small scale spurious motions we then employed the regularized displacement fields with transfer functions defined in \cite{Baldauf:2015tla}. 

We find that in our highest order calculation, at redshift $z=0$ the power spectrum of the density field is reproduced with an accuracy of $1\%$ ($10\%$)  up to $k=0.25\ihMpc$ ($k=0.46\ihMpc$). We believe that the dominant source of the remaining error is the stochastic contribution, orthogonal to the perturbative basis. 
The stochastic term will likely put an upper limit on the range over which the non-linear power spectrum can be modeled by IR-resummed two-loop Eulerian EFT. Perturbation theory approaches should generally only aim to model the deterministic (non-stochastic) part of the density field.

The Eulerian stochastic term deviates from the Lagrangian stochastic term of the displacement divergence on all but the largest scales. This is expected from the collapse of structure and can be explained quantitatively by the three point function of the stochastic term itself and correlations between the square of the stochastic term and perturbation theory.
The stochastic term only scales as $k^4$ on the largest scales, being shallower in the range where it starts to affect the power spectrum at percent level and finally asymptoting to the halo profile. These deviations from the simple $k^4$ behaviour over the relevant scales will greatly complicate the modelling of the stochastic term, requiring additional parameters or scales besides the non-linear scale.

\section*{Acknowledgements}

The authors would like to thank Mehrdad Mirbabayi, Uro\v{s} Seljak, Leonardo Senatore, Marko Simonovi\'{c}, Zvonimir Vlah and Martin White for fruitful discussions.
T.B. is supported by the Institute for Advanced Study through a Corning Glass Works foundation fellowship.
E.S. is supported by the NSF grant AST1311756 and the NASA grant NNX12AG72G.
M.Z. is supported in part by the NSF grants  PHY-1213563 and AST-1409709. 
While this paper was being finished, a similar study \cite{Vlah:2015sea} was published. We reach similar conclusions to this study where there is overlap.

\bibliographystyle{JHEP}
\bibliography{lagtoeul}

\appendix
\section{Cross checks}\label{sec:crosscheck}

\begin{figure}
\centering
\includegraphics[width=0.49\textwidth]{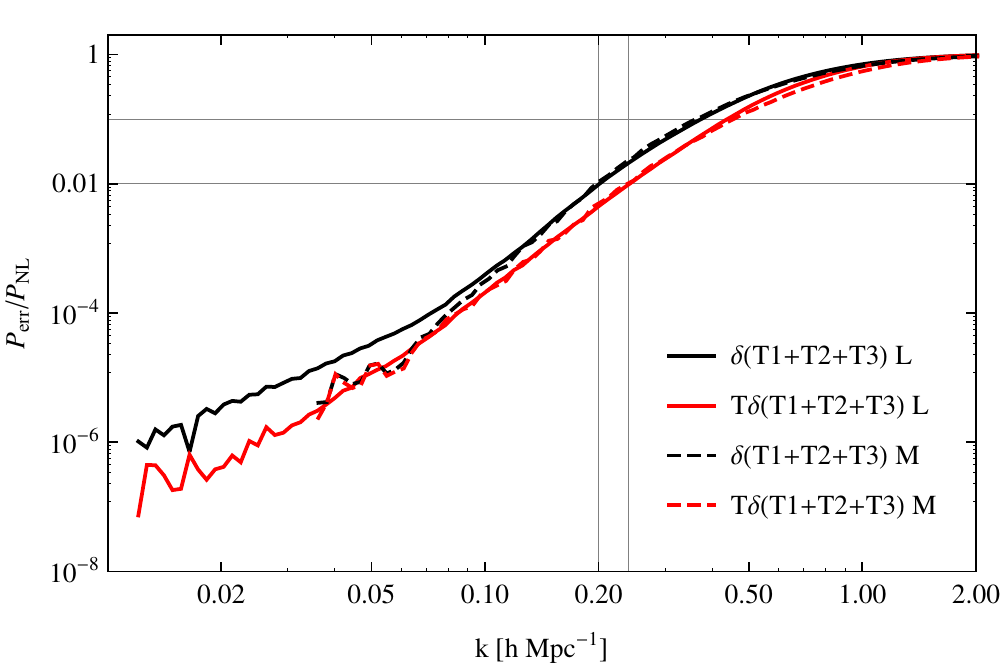}
\includegraphics[width=0.49\textwidth]{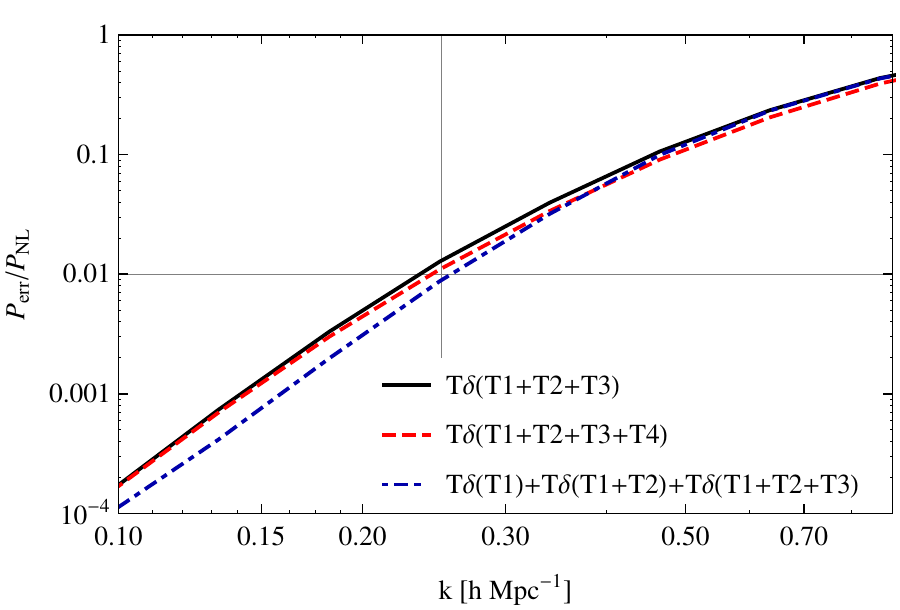}
\caption{
\emph{Left panel: }Comparison between ratio of error and non-linear power spectra for the M and L simulations before and after employing the transfer functions. Despite the very different box size and cutoff wavenumbers of the LPT calculation, we find extremely good agreement between the error power spectra.
\emph{Right panel: }Comparison of the error power spectra in the M simulation with three transfer functions on the displacement fields and an overall transfer function on the density field with the cases where we add the fourth order displacement with the corresponding transfer function and the case where the density fields generated from the first, second and third order displacements have different density transfer functions.}
\label{fig:errorLvsM}
\label{fig:multipleTs}
\end{figure}

In the left panel of Fig.~\ref{fig:errorLvsM} we compare the error power spectra between the L and M runs. Despite the varying simulation resolution and the difference in the cutoffs employed for the perturbative calculation the results are very well converged on the scales of interest. In particular, the scale where the stochastic term corresponds to one percent of the non-linear power spectrum agrees between the two cases. We can thus conclude that we have employed sufficiently general transfer functions to capture the cutoff dependence of the theory.

In the right panel of Fig.~\ref{fig:multipleTs} we show the error power spectrum for a few cases not explicitly considered in the main text in order not to confuse the reader. We start from our fiducial model $T\delta(T1+T2+T3)$ and generalize it with additional transfer functions. As a first approach we add the fourth order displacement field in Lagrangian space multiplied with its transfer function. Considering the error power spectrum arising from the corresponding density field with an additional density transfer function, we see no considerable improvement over the fiducial case. As the most general case we split the basis for the density field, considering the basis vectors $\delta(T1)$, $\delta(T1+T2)$ and $\delta(T1+T2+T3)$ separately and allowing three separate transfer functions for these density fields. This model has in total six free functions of wavenumber (three in Lagrangian and three in Eulerian space). Even in this fairly general approach, the improvement in the $k$-reach is negligible. However, one very large scales the error at a fixed $k$ decreases by about a factor of two. 
For the latter term we can explicitly show, that it should contain all the terms in two loop SPT and their counterterms. The 1-5 contribution is accounted for by a transfer function on $\delta(T1)$, the 2-4 contribution is accounted for by a transfer function on the $\delta(T1+T2)-\delta(T1)$ and the 3-3 contribution is accounted for by $\delta(T1+T2+T3)-\delta(T1+T2)$ for which we actually wouldn't even need a transfer function to get to two loops. Concerning the counterterms, they come in the form of $\tilde 1 - 1$, $\tilde 2 - 2$, $\tilde 3 - 1$ and $\tilde 1 - \tilde 1$ (where the tilde stands for a counterterm involving density fields of this order). All of these terms have been included by the transfer functions that we are allowing for.

\section{Equivalence with 2-loop SPT}\label{app:equiv}

When expanded SPT and Lagrangian solutions are equivalent. This means that
\be
\delta^{(\bar 5)}=\delta^{(1)}+\delta^{(2)}+\delta^{(3)}+\delta^{(4)}+\delta^{(5)}+ \cdots
\ee
where the dots stand for terms that are higher order than what we need at two loops. We can now rewrite:
\bea
\label{barexp}
\delta^{(\bar 5)}&=&(\delta^{(\bar 5)}-\delta^{(\bar 4)})+(\delta^{(\bar 4)}-\delta^{(\bar 3)})+(\delta^{(\bar 3)}-\delta^{(\bar 2)})+(\delta^{(\bar 2)}-\delta^{(\bar 1)})+\delta^{(\bar 1)}, \nonumber \\
&=& \delta^{(\hat 5)} + \delta^{(\hat 4)} + \delta^{(\hat 3)}+\delta^{(\hat 2)}+\delta^{(\hat 1)}
\eea
with the definition $\delta^{(\hat n)} \equiv \delta^{(\bar n)}-\delta^{(\bar n-1)}$. We note that $\delta^{(\hat n)}$ starts at order $n$, but includes higher orders. In fact it correctly includes the effects of large scale bulk motions and this is why it is the convenient basis to compare directly with the density of $N$-body simulations by means of cross correlations to determine transfer functions (or EFT counter terms). 

Because of the equivalence between SPT and the Lagrangian treatment when expanded we also have:
\be
P_{\bar 5\bar 5} = P_\text{2-loop, E} + \cdots
\ee
where again the dots stand for higher order terms. Now we can use equation (\ref{barexp}) to write:
\be
P_{\bar 5\bar 5} = P_{\hat 1\hat 1} + P_{\hat 2\hat 2}+ P_{\hat 3\hat 3} + 2 P_{\hat 1\hat 2} + 2 P_{\hat 1\hat 3}+ 2 P_{\hat 1\hat 4} + P_{\hat 1\hat 5} + 2 P_{\hat 2\hat 3} + 2 P_{\hat 2\hat 4} + \cdots
\ee
where we have left out any higher order terms, such as $P_{\hat 4\hat 4}$.  
We now see that the only parts of  $\delta^{(\hat 4)}$ and $\delta^{(\hat 5)}$ that are needed for this calculation are those that correlate with $\delta^{(\hat 1)}$ or $\delta^{(\hat 2)}$. Thus all that is needed at the level of the fields is the basis  $(\delta^{(\hat 1)},\delta^{(\hat 2)},\delta^{(\hat 3)})$. So at the level of the fields and at two loop order we could write 
\be
\delta^{(\bar 5)} = T_{(\hat 1)}  \delta^{(\hat 1)} + T_{(\hat 2)} \delta^{(\hat 2)}+ T_{(\hat 3)} \delta^{(\hat 3)} = T_{(\bar 1)}  \delta^{(\bar 1)} + T_{(\bar 2)} \delta^{(\bar 2)}+ T_{(\bar 3)} \delta^{(\bar 3)}
\ee
to recover all the terms that enter in a two loop calculation. In the last line we reexpressed $\delta^{(\hat n)}$ in terms of $\delta^{(\bar n)}$. Besides the terms in the SPT two loop calculation the transfer functions will also pick up the counterterms to the extend that they are present in the data and correlate with first, second and third order fields.

\end{document}